\documentclass[journal,10pt,onecolumn]{IEEEtran}
\usepackage{graphicx} 
\usepackage{amsmath} 
\usepackage{amssymb}
\usepackage{cite}
\usepackage{amsfonts}
\usepackage{amssymb}
\usepackage{caption}
\usepackage{subcaption}
\usepackage{psfrag}
\usepackage{array}
\usepackage{multirow}
\usepackage{multicol}
\usepackage{setspace}
\usepackage{longtable}
\usepackage{float}
\usepackage{color}
\usepackage[table]{xcolor}

\newcommand{\bm}{\mathbf}
\newcommand{\be}{\begin{equation}}
\newcommand{\ee}{\end{equation}}
\newcommand{\bea}{\begin{eqnarray}}
\newcommand{\eea}{\end{eqnarray}}

\newcommand{\ba}{{\bm a}}

\newcommand{\dd}{{\bm d}}
\newcommand{\bA}{{\bm A}}
\newcommand{\bI}{{\bm I}}
\newcommand{\bW}{{\bm W}}
\newcommand{\bE}{{\bf E}}
\newcommand{\bF}{{\bf F}}
\newcommand{\bD}{{\bf D}}
\newcommand{\bC}{{\bf C}}

\newcommand{\bS}{{\bf S}}
\newcommand{\bH}{{\bf H}}
\newcommand{\bP}{{\bf P}}

\newcommand{\bg}{{\bf g}}
\newcommand{\by}{{\bf y}}

\newcommand{\bh}{{\bf h}}

\newcommand{\bV}{{\bf V}}
\newcommand{\bU}{{\bf U}}

\newcommand{\bd}{{\bf d}}
\newcommand{\dt}{{\tilde{d}}}

\newcommand{\bx}{{\bf x}}

\newcommand{\bnu}{\mbox{\boldmath$\nu$}}

\newcommand{\BG}{{\boldsymbol{\mathcal G}}}

\newcommand{\mmseB}{{\boldsymbol{\mathcal B}}}
\newcommand{\mmseC}{{\boldsymbol{\mathcal C}}}

\newcommand{\BLambda}{\mbox{\boldmath$\Lambda$}}

\newcommand{\AHA}{\mbox{$\bA^{\rm H}\bA$}}

\newcommand{\ANHA}{\mbox{$\bA_N^{\rm H}\bA_N$}}
\newcommand{\HAHA}{\mbox{$(\bH\bA)^{\rm H}\bH \bA$}}
\newcommand{\HA}{\mbox{$(\bH\bA)$}}

\newcommand{\mI}{\mbox{$\frac{\sigma_\nu^2}{\sigma_d^2}\bI$}}
\newcommand{\snri}{\mbox{$\frac{\sigma_\nu^2}{\sigma_d^2}$}}
\newcommand*{\Scale}[2][4]{\scalebox{#1}{$#2$}}%
\newcommand{\NFD}{\mbox{\boldmath{$\Psi$}}}

\newcommand\undermat[2]{%
  \makebox[0pt][l]{$\smash{\underbrace{\phantom{%
    \begin{matrix}#2\end{matrix}}}_{\text{$#1$}}}$}#2}

\author{Shashank~Tiwari,~Suvra~Sekhar~Das and~Kalyan~Kumar~Bandyopadhyay
\thanks{Authors are with the Indian Institute of Technology, Kharagpur. email:~{shashank,suvra}@gssst.iitkgp.ernet.in,~kalyan@ece.iitkgp.ernet.in}}

\begin{document}
\bstctlcite{IEEEexample:BSTcontrol}
\title{Precoded GFDM System to Combat Inter Carrier Interference : Performance Analysis }
\maketitle

\begin{abstract}
The expected operating scenarios of 5G pose a great challenge to orthogonal frequency division multiplexing (OFDM) which has poor out of band (OoB) spectral properties, stringent synchronization requirements, and large symbol duration. Generalized frequency division multiplexing (GFDM) which is the focus of this work, has been suggested in the literature as one of the possible solutions to meet 5G requirements. In this work, the analytical performance evaluation of MMSE receiver for GFDM is presented. We also proposed precoding techniques to enhance the performance of GFDM. A simplified expression of SINR for MMSE receiver of GFDM is derived using special properties related to the modulation matrix of GFDM, which are described in this work. This SINR is used to evaluate the BER performance. Precoding schemes are proposed to reduce complexity of GFDM-MMSE receiver without compromising on the performance. Block Inverse Discrete Fourier Transform (BIDFT) and Discrete Fourier Transform (DFT) based precoding schemes are found to outperform GFDM-MMSE receiver due to frequency diversity gain while having complexity similar to zero-forcing receiver of GFDM. It is  shown that both BIDFT and DFT-based precoding schemes reduce peak to average power ratio (PAPR) significantly. Computational complexity of different transmitters and receivers of precoded and uncoded GFDM is also presented.
\end{abstract}

\IEEEpeerreviewmaketitle
\section{Introduction} \label{sec:introduction}
The need for wireless links, from proximity to a large distance communication, has been on the rise since its origin without ever showing any signs of distaste to growth. Towards this, several applications with diverse needs have driven the development of technical solutions. Amongst many, the domain of public wireless communication has provided one of the greatest benefits and drive to the development of society. The success is powered by standardization, which led to mass production and adoption of technology due to cost effectiveness.
The development of 1G - 4G \cite{sesia,holmalte} have fundamental drivers as the \textit{need for speed} (higher data rate and higher spectral efficiency) and the requirement of lower latency. The next generation, 5G is also expected to be driven by similar demands but with added needs such as flexibility. Some of the important applications operating 5G \cite{wunder_5gnow:_2013,metis_general,santhi_goals_2003,tactile} are flexible radio (cognitive radio), tactile internet, machine type communication (MTC), great service in crowd, super real time services, super reliable communication, and others.
OFDM is de-facto transmission technology for broadband wireless access due to the robustness to multipath fading, high spectral efficiency and ease of implementation but has limited capability in meeting the needs of new requirements of 5G \cite{wunder145gnow,banelli14modulation}. MTC demands relaxed synchronization requirement which is a limitation in OFDM as it is sensitive to frequency errors. Small symbol duration is needed for low latency applications and due to the cyclic prefix (CP), OFDM will lead to low efficiency. High OoB radiation of OFDM limits, it's use for opportunistic use of spectrum and dynamic spectrum allocation.\\
\indent To meet above requirements, Filter Bank Multi Carrier (FBMC)\cite{bellanger_physical_2010,FMT,SMT,CMT}, Constant Envelope-OFDM (CE-OFDM) \cite{swindlehurst_millimeter-wave_2014}, GFDM \cite{michailow_generalized_2012}, Unified Filtered Multi Carrier (UFMC) \cite{vakilian_universal-filtered_2013} have been proposed, where each has it's own special features.
FBMC, whose concept dates back to 1967 \cite{chang66}, is currently being considered as a candidate for 5G waveform due to it's good frequency localization capabilities as well as low OoB radiation. However, FBMC signal suffer from inter symbol interference (ISI) as no cyclic prefix (CP) is used. The FBMC signal stretches in time and hence, limits spectral efficiency gains. Hence, FBMC is not suitable for low latency applications \cite{schaich_waveform_2014}. UFMC (proposed for low latency applications) has limited suppression of OoB radiation. In this work, we focus on GFDM, which is another competitive waveform. GFDM has been shown to be flexible \cite{fettweis_gfdm_2009,wunder_5gnow:_2013, GFDM_tr} in terms of using time-frequency resources. It has good OoB radiation properties \cite{matthe_influence_2014}. It is also quite resilient to synchronization requirement \cite{GFDM_tr} and has good spectral efficiency as it uses circular pulse shaping which reduces cyclic prefix length in frequency selective fading channel(FSFC).\\
\indent Although GFDM has attractive features, yet there remains sufficient investigations to be done before it can be adopted into practical communication systems. One of the fundamental differences between GFDM and OFDM is that the former uses non-rectangular pulse shape on each subcarrier where it was rectangular for the latter. This causes inter-carrier interference (ICI) \cite{gaspar_low_2013}. Therefore, receiver design is an important issue in GFDM. It was earlier quite simple in the case of OFDM which led to its mass adoption. GFDM receiver in FSFC has been proposed as a two-stage receiver \cite{GFDM_ber_fs}. Frequency domain zero forcing (ZF) equalization is used to mitigate the effect of the wireless channel in the first stage \cite{GFDM_tr}. In second stage, linear as well as non-linear receivers have been proposed to mitigate self-interference in GFDM systems\cite{michailow_bit_2012, datta_gfdm_2012}. Linear receivers of GFDM such as Matched filter (MF), Zero Forcing (ZF) and Minimum Mean Square Error (MMSE) are proposed in \cite{michailow_bit_2012}. It has been shown that MF is not able to mitigate self-interference. It has also been shown in \cite{michailow_bit_2012, GFDM_ber_fs} that MMSE performs better than ZF, especially when signal to interference and noise ratio(SINR) is low. For high SINR, MMSE performance matches with ZF. Analytical expression for bit error rate (BER) for MF and ZF receiver over additive white Gaussian noise channel (AWGN) is given in \cite{GFDM_tr,matthe_influence_2014}. Analytical expression of SINR is an essential part of performance analysis of a system as it leads to the calculation of error probability and capacity. BER computation for MMSE receiver is still not available in the literature.\\
\indent Since implementation of MMSE receiver requires very high amount of complexity, Successive Interference Cancellation (SIC) receivers to mitigate self-interference in AWGN channel were proposed in\cite{datta_gfdm_2012}. Double SIC (D-SIC) receiver introduced in \cite{datta_gfdm_2012} cancels interference from two adjacent subcarriers of all symbols. D-SIC can cancel out self-interference entirely in AWGN channel but requires few iterations thus induces processing delay at the receiver. For usability of GFDM, low complexity signal processing techniques remains to be investigated. \\
\indent Peak to Average Power Ratio (PAPR) is an important issue in the multicarrier communication.We expect GFDM to have a higher PAPR than OFDM as the use of identical non-rectangular pulse shapes increase PAPR in multicarrier system \cite{slimane_peak--average_2000}. Authors in \cite{michailow_low_2013} have compared PAPR of GFDM with OFDM for unequal number of subcarriers. As far as we know, PAPR comparison of GFDM with OFDM for equal number of subcarriers has not been investigated. Apart from that PAPR reduction schemes also need to be studied.\\
\indent A detailed exposition of the product of the modulation matrix with its hermitian reveals some interesting properties which helps in performance analysis and developing precoding techniques. The details are described in this paper. A simplified expression for SINR of MMSE receiver is developed using the above mentioned special properties. It is also shown that interference plus noise values can be approximated as a Gaussian random variable. Analytical BER is computed in AWGN and FSFC using derived SINR expressions.\\
\indent A detailed complexity analysis of different schemes is presented in this paper which shows that D-SIC is not quite simple to implement in comparison to ZF receiver. To reduce the self-interference at the receiver, precoding techniques for GFDM are investigated in this work. A generalized framework for precoding based GFDM is developed. Based on the properties related to the modulation matrix as mentioned above, Block Inverse Discrete Fourier Transform (BIDFT) based precoding is proposed. Performance of BIDFT, Discrete Fourier Transform (DFT) and  Singular Value Decomposition (SVD) based precoding techniques are compared with uncoded GFDM. PAPR of precoded GFDM is compared with uncoded GFDM and OFDM as well.\\
\indent The rest of the paper is organized as follows. The system model is developed in Section~\ref{gfdm:ietcomm:system_model}.
Analytical BER for MMSE receiver is presented in Section~\ref{wpc'14:gfdm:ber:mmse}. The precoding schemes proposed in this work are presented in Section~\ref{sec:gfdm:ietcomm:precoding:precoding}. Results related to the proposed works are given in Section~\ref{sec:result}. Section~\ref{sec:gfdm:ietcomm:precoding:conclusion} has the conclusion. \\
\indent In this work vectors are represented by bold small letters, matrices are represented by bold capital letters and scalers are represented as normal small letters. The operation $a(n ~{\texttt{modulus}}~ l)$ is written as $a(n)_{l}$, $\bI_N$ represents identity matrix with order $N$ and `$\ast$' represents convolution operation. $E$ is expectation operation and $j=\sqrt{-1}$.

\section{System Model} \label{gfdm:ietcomm:system_model}
\begin{figure}[h]
\psfrag{AA}[L]{ $\dd$}
\psfrag{BB}[R]{ \small{$(\bP)$}}
\psfrag{CC}[L]{ $\small{\tilde{\dd}}$}
\psfrag{TT}[B]{\small{$(\bA)$}}
\psfrag{EE}{$\bx$}
\psfrag{FF}[B][[C]{$\bx_{\texttt{CP}}$}
\psfrag{SS} {\small{$(\bH$})}
\psfrag{HH} [B]{\tiny{$\bH \bA \tilde{\dd}$}}
\psfrag{GG} [t]{\small{$\bnu$}}
\psfrag{II} [B]{\tiny{$(\bH^{\rm -1})$}}
\psfrag{JJ} [L]{\tiny{$\by_{\texttt{FDE}}$}}
\psfrag{KK} [B]{\tiny{$({\bD})$}}
\psfrag{NN} [R]{\tiny{$(\bC)$}}
\psfrag{OO} [R]{{$\hat{\tilde{\dd}}$}}
\psfrag{RR} [B]{\hspace{0.5cm}{$\hat{{\dd}}$}}
\psfrag{MM} [B]{{\tiny{$(\bP^{\rm H})$}}}
\psfrag{PP} [B]{{\tiny{$(\bP^{\rm H})$}}}
\psfrag{UU} [B] {}
\psfrag{Equation}[L] {\Scale [0.6]{\hspace{1cm}\bf{Receiver}}}
\includegraphics[trim=0cm 2cm 0cm 0cm, width = \linewidth]{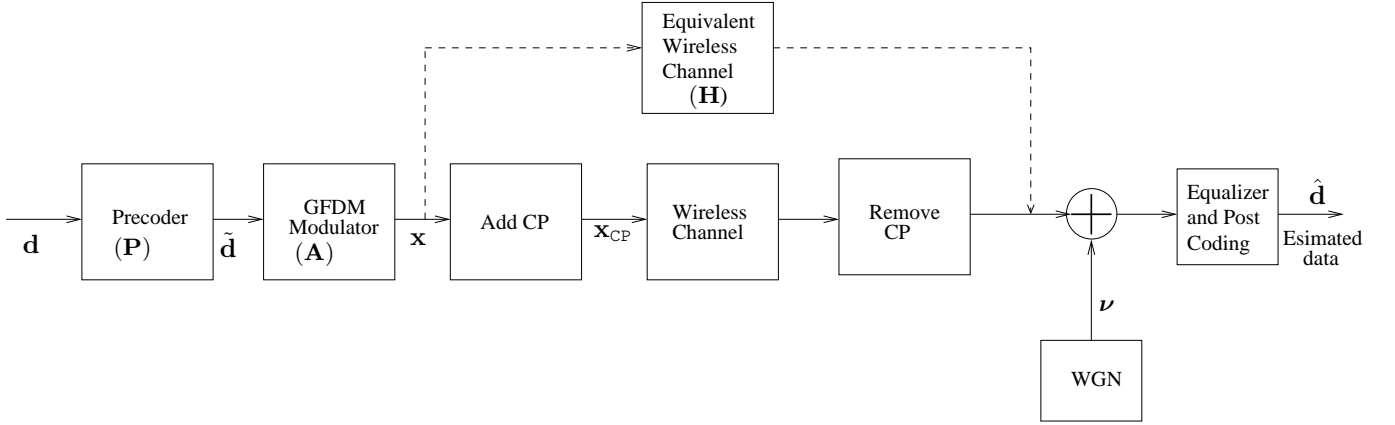}
\vspace{-4cm}
\caption{Transmitter and receiver architecture for Precoding based GFDM system}
\label{fig:gfdm:letter:precoding:precodingTxRx}
\end{figure}

\indent GFDM is a multi carrier modulation technique with some similarity to OFDM. We begin by considering a block of QAM modulated symbols $\dd = [d_0 ~ d_1\cdots d_{MN-1}]^{\rm T}$. We assume that data symbols are independent and identical i.e. $E[d_l d_l^\ast]= \sigma_d^2 $ $, \forall l$ and $E[d_l d_{q}^\ast]= 0 $ when $l\neq q$. Let the total bandwidth $B$ be divided into $N$ number of sub-carriers where symbol duration $T = \frac{N}{B}$ second and $\frac{B}{N}$ Hz is the subcarrier bandwidth. In case of OFDM this leads to orthogonal subcarriers. Let the symbol duration, $T$, be one time slot. GFDM is a block based transmission scheme and we consider a block to have $M$ such time slots. Hence, in one block there are $N ~\text{subcarriers}~ \times M~\text{timeslots}~= NM$ QAM symbols.
The flow of operations, as described below, can be understood in the light of Fig.~\ref{fig:gfdm:letter:precoding:precodingTxRx}.
\subsection{Transmitter}
\begin{figure}[h]
\includegraphics[width=\linewidth]{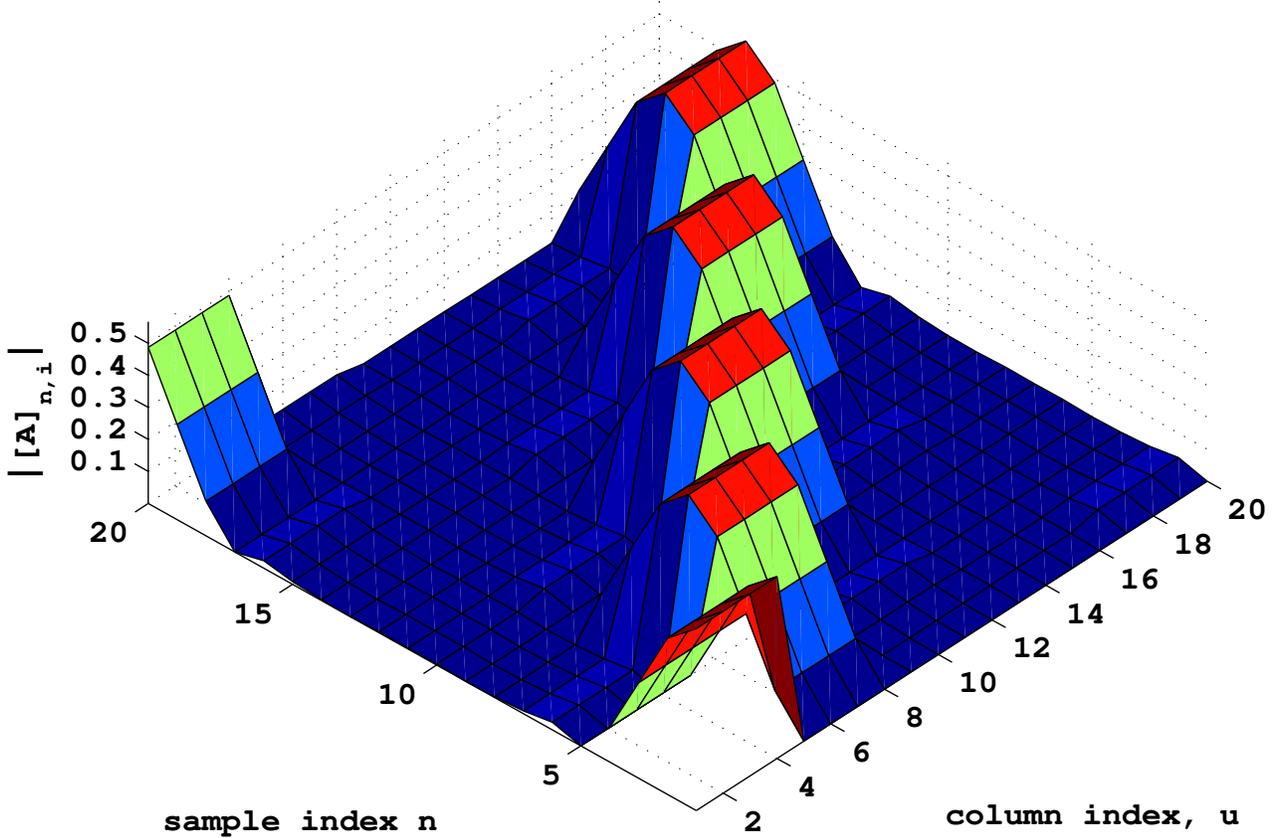}
\caption{Time Domain view of columns of $\bA$ for $N =4$, $M=5$ and $ROF =0.9$}
\label{fig:GFDM:A:matrix:time}
\end{figure}

Since we consider precoding, let $\bP$ be a precoding matrix of size $MN \times MN$, which is defined in Section~\ref{sec:gfdm:ietcomm:precoding:precoding}. The data vector $\dd$ be multiplied with precoding matrix $\bP$ and we obtain precoded data vector $\tilde{\dd} = \bP \dd$. The $MN \times 1$ precoded data vector $\tilde{\dd} = [\tilde{d}_{0,0} \cdots \tilde{d}_{k,m} \cdots \tilde{d}_{N-1,M-1}]^{\rm T}$, where $k= 0 \cdots N-1$ denote subcarrier index and $m= 0 \cdots M-1$ indicates time slot index. Conventional GFDM system \cite{GFDM_tr} can be seen as a special case of precoded GFDM system when, $\bP = \bI_{NM}$. The precoded data vector, $\tilde{\dd}$, is modulated using GFDM Modulator.
Precoded data $\tilde{d}_{k,m}$ is first upsampled by $N$, which is represented as,
\be
\tilde{d}^{up}_{k,m}(n) = \tilde{d}_{k,m} \delta(n-mN),~ n=0,1, \cdots , MN-1.
\ee
Now, this upsampled data is pulse shaped. Impulse response of pulse shaping filter is represented by $g(n)$. Its length is $MN$. The upsampled precoded data $\tilde{d}^{up}_{k,m(n)}$ is circularly convoluted with such pulse shaping filter $g(n)$ and can be written as,
\be
\begin{aligned}
x^{f}_{k,m}(n) &= \sum_{r=0}^{MN-1}{\tilde{d}_{k,m}g(n-r)_{MN}\delta(r-mN)}
&= \tilde{d}_{k,m}g(n-mN)_{MN}.
\end{aligned}
\ee
Now the filtered data is up-converted to $k^{\rm th}$ subcarrier frequency and is given as,
\be
\begin{aligned} \label{eqn:x_km}
x_{k,m}(n) &= \tilde{d}_{k,m}g(n-mN)_{MN} e^{\frac{j2\pi k n}{N}} &= \tilde{d}_{k,m} a_{k,m}(n),
\end{aligned}
\ee
where, $a_{k,m}(n)=g(n-mN)_{MN} e^{\frac{j2\pi k n}{N}}$ for $n=0,~1, \cdots MN-1$, is called the kernel of GFDM for $k^{\rm th}$ subcarrier and $m^{\rm th}$ time slot. Now, output samples of all subcarrier frequencies and time slots are added and the GFDM signal can be given as,
\be \label{eqn:x}
\begin{aligned}
x(n) &= \sum_{m=0}^{M-1}{\sum_{k=0}^{N-1}{x_{k,m}(n)}}
&= \sum_{m=0}^{M-1}{ \sum_{k=0}^{N-1}{\tilde{d}_{k,m} } a_{k,m}(n)} .
\end{aligned}
\ee
If we let, $l= mN+k$ for $m=0,~1 \cdots M-1$ and $k=0,~1 \cdots N-1$. Then, we may write (\ref{eqn:x}) as,

\be \label{eqn:x:l}
\begin{aligned}
x(n) &= \sum_{l=0}^{MN-1}{ a_l^N(n) \dt_l^N} ,
\end{aligned}
\ee
where, superscript $N$ is identifier of the specific mapping between index $l$ and time slot and subcarrier index tuple $(k,m)$ which takes $N$ as scaler multiplier in the mapping. If we collect all output samples of GFDM signal in a vector called $\bx=[x(0) ~x(1) \cdots x(MN-1)]^{\rm T}$ and all samples of $a_l^N(n)$ is a vector called $\ba_l^N= [a_l^N(0)~a_l^N(1) \cdots a_l^N(MN-1)]$. Then using (\ref{eqn:x:l}), $\bx$ can be written as,
\be \label{eqn:x:vector}
\begin{aligned}
\bx &= \sum_{l=0}^{MN-1}{ \ba_l^N \dt_l^N} &\equiv \sum_{l=0}^{MN-1}{\dt_l^N \times~\text{$l^{\rm th}$ column vector}~\ba_l^N}
&= \bA_N \tilde{\bd}^N,
\end{aligned}
\ee
where, $\bA_N = [ \ba^N_0 ~\ba^N_1 \cdots \ba^N_{MN-1}]$ is modulation matrix , $\tilde{\bd}^N = [\dt_0 ~ \dt_1 \cdots \dt_l \cdots \dt_{NM-1}]^{\rm T}$ is precoded data vector. If we collect all samples of $a_{k,m}(n)$ is vector called $\ba_{k,m}= [a_{k,m}(0)~a_{k,m}(0)\cdots a_{k,m}(MN-1)]^{\rm T}$, then $\bA_N$ can also be written as,
\small \be \label{eqn:A_N}
\begin{aligned}
\bA_{\rm N} &= [\underbrace{\overbrace{\ba_{0,0}}^{\substack{\texttt{1}^{st}\\{\texttt{freq.}}}}~
\overbrace{\ba_{1,0}}^{\substack{\texttt{2}^{nd}\\{\texttt{freq.}}}} \cdots \overbrace{\ba_{N-1,0}}^{\substack{\texttt{N}^{th}\\{\texttt{freq.}}}}}_{1^{\rm st} \texttt{time slot}} \vert \underbrace{\ba_{0,1}~ \ba_{1,1} \cdots \ba_{N-1,1}}_{2^{\rm nd} \texttt{time slot}} \vert \cdots \vert \underbrace{\ba_{0,M-1}~ \ba_{1,M-1} \cdots \ba_{N-1,M-1}}_{M^{\rm th} \texttt{time slot}}] \\ \vspace{0.5cm}
&=\Scale[0.7]{{\begin{bmatrix}
g(0)&g(0) & \cdots & \cdots & g(0) & g(\texttt{MN-N}) & \cdots & \cdots & g(\texttt{MN-N})& \cdots\\
g(1)&g(1)\rho & \cdots & \cdots & g(1)\rho^{\texttt{N-1}}& g(\texttt{MN-N+1}) & \cdots &\cdots & g(\texttt{MN-N+1}) \rho^{\texttt{N}-1}&\vdots \\
\vdots & \vdots & \cdots &\cdots & \vdots & \vdots & \cdots &\cdots & \cdots & \texttt{(M-2)N terms} \\
\vdots & \vdots & \cdots & \cdots & \vdots & \vdots & \cdots & \cdots & \cdots & \vdots \\
\undermat{\ba_{\rm 0,0}}{g(\texttt {MN-1})}&\undermat{\ba_{\rm 1,0}}{g(\texttt {MN-1})\rho^{\texttt {MN-1}}} & \cdots & \cdots & \undermat{\ba_{\texttt{ N-1,0}}}{g(\texttt{MN-1})\rho^{\texttt{(N-1)(MN-1)}}}&\undermat{\ba_{\rm 0,1}} {g(\texttt{MN-N-1})} & \cdots & \cdots & \undermat{\ba_{\texttt{N-1,1}}}{g(\texttt{MN-N-1}) \rho^{\texttt{(N-1)(MN-1)}}} & \cdots\\
\end{bmatrix}}},
\end{aligned}
\vspace{0.5cm}
\ee
\normalsize
where, $\rho=e^{\frac{j2\pi}{N}}$. At this point, it is also interesting to look into the structure $\ba_{k,m}$ vectors which constitute columns of $\bA_N$. The first column of $\bA_N$ i.e. $\ba_{0,0}$ holds all coefficient of pulse shaping filter and other columns of $\bA_N$ or other vectors in the set of $\ba_{k,m}$'s are time and frequency shifted version of $\ba_{0,0}$ where frequency index $k$ denotes $\frac{k}{N}$ shift in frequency and time index $m$ denotes $m$ time slot or $mN$ sample cyclic shift.
Taking clue from description of column vectors $\ba_{k,m}$, structure of $\bA_N$ can be understood by (\ref{eqn:A_N}). Columns having same time shift are put together. Columns which are $N$ column index apart are time shifted version of each other. Columns having same time shift are arranged in increasing order of frequency shift. We will explore time domain and frequency domain behaviour for an example $\bA_N$ with total subcarriers $N=4$ and total time slots $M=5$. Pulse shaping filter is taken to be root raised cosine (RRC) with roll of factor (ROF) of 0.9. Fig.~\ref{fig:GFDM:A:matrix:time} shows the absolute values of each column index $u=0,~1\cdots MN-1$ with sample index $n= 0,~1\cdots, MN-1$. It can be observed that first $N=4$ columns have $0$ time shift and next $N$ columns have unit time shift and so on.
\newline Alternatively, if we take $l=kM+m$ in equation (\ref{eqn:x}), modulation matrix $\bA_{\rm M}$ can be represented as,
\small \be \label{eqn:A_M}
\bA_{\rm M} = [\underbrace{\overbrace{\ba_{0,0}}^{\substack{\texttt{1}^{st}\\{\texttt{time}}}}~
\overbrace{\ba_{0,1}}^{\substack{\texttt{2}^{nd}\\{\texttt{time}}}} \cdots \overbrace{\ba_{0,M-1}}^{\substack{\texttt{M}^{th}\\{\texttt{time}}}}}_{1^{\rm st} \texttt{frequency}} \vert \underbrace{\ba_{1,0}~ \ba_{1,1} \cdots \ba_{1,M-1}}_{2^{\rm nd} \texttt{frequency}} \vert \cdots \vert \underbrace{\ba_{N-1,0}~ \ba_{N-1,1} \cdots \ba_{N-1,M-1}}_{N^{\rm th} \texttt{frequency}}].
\ee
\normalsize
Matrix $\bA_N$ and $\bA_M$ can be related as $\bA_N = \boldsymbol{\zeta}\bA_M$, where $\boldsymbol{\zeta}$ is a permutation matrix which permutes column of matrix applied.
Cyclic prefix is added to GFDM modulated block to prevent inter-block interference in FSFC. CP of length $N_{CP}$ is prepended to $\bx$. After adding CP, transmitted vector, $\bx_{cp}$, can be given as
\be \label{x_cp}
\bx_{cp} = [\bx(MN-N_{cp}+1 : MN)\hspace{0.2 cm} ;~ \bx]
\ee
In the rest of the paper, for equations which are valid for both $\bA_M$ and $\bA_N$, modulation matrix will be denoted as $\bA$. However, wherever required $\bA_N$ and $\bA_M$ will be specified.
 
 \subsection{Receiver}
 Let, $\bh=[h_1,~h_2,\cdots h_{L}]^{\rm T}$ be $L$ length  channel impulse response vector, where, $h_r,~\text{for}~1 \leq r \leq L$, represents the complex baseband channel coefficient of $r^{\rm th}$ path \cite{proakis}, which we assume is zero mean circular symmetric complex Gaussian (ZMCSC). We also assume that channel coefficients related to different paths are uncorrelated. We consider, $N_{cp}\geq L$. Received vector of length $N_{CP}+NM+L-1$ is given by,
\be \label{y_cp}
\by_{cp} = \bh \ast \bx_{cp} + \bnu_{cp},
\ee
where $\bnu_{cp}$ is AWGN vector of length $MN+N_{cp}+L-1$ with elemental variance $\sigma_\nu^2$. 
 The first $N_{cp}$ samples and last $L-1$ samples of $\by_{cp}$ are removed at the receiver i.e. $\by = [\by_{cp}(N_{cp}+1 : N_{cp}+MN)]$. Use of cyclic prefix converts linear channel convolution to circular channel convolution when $N_{cp}\geq L$\cite{sesia}. The $MN$ length received vector after removal of CP can be written as,
  \be \label{eqn: rec}
  \begin{aligned}
  \by &= \bH \bA \tilde{\dd} + \bnu,
  \end{aligned}
  \ee
  where $\bH$ is  circulant convolution matrix of size $MN \times MN$, which can be written as,
\begin{equation}
 \bH = \begin{bmatrix}
  h_1 & 0 &\cdots& 0 & h_{L} & \cdots & h_{2}\\
  h_2 & h_1 & \cdots & 0 & 0 & \cdots& h_{3}\\
  \vdots& & \ddots \\
  h_{L} & h_{L-1} & \cdots & \cdots & \cdots &\cdots & 0\\
  0 & h_{L} & \cdots & \cdots & \cdots &\cdots & 0\\
  \vdots& & \ddots \\
  0 & 0 & &h_{L} & \cdots& \cdots& h_1
 \end{bmatrix},
\end{equation}
and $\bnu$ is WGN vector of length $MN$ with elemental variance $\sigma_\nu^2$. 
Received vector $\by$ is distorted due to (i) Self interference as subcarriers are non-orthogonal\cite{michailow_bit_2012} and (ii) Inter-carrier interference(ICI) due to FSFC.
 
\section{Bit Error Rate Computation  For MMSE Receiver} \label{wpc'14:gfdm:ber:mmse}
\subsection{MMSE Receiver}
To equalize the channel and GFDM induced self interference, a joint MMSE equalizer \cite{GFDM_tr} is considered here. Equalized data can be given as,
\be \label{eqn:mmse:rx}
\begin{aligned}
\hat{\dd} &= [\bI_{MN}\frac{\sigma_{\nu^2}}{\sigma_{d^2}}+ (\bH \bA)^{\rm H} (\bH \bA)]^{-1} (\bH \bA)^{\rm H} \by\\
&= \mmseB \dd + \mmseC \bnu ,
\end{aligned}
\ee
where, $\mmseB = [\bI_{MN}\frac{\sigma_{\nu^2}}{\sigma_{d^2}}+ (\bH \bA)^{\rm H} (\bH \bA)]^{-1} (\bH \bA)^{\rm H} (\bH \bA)$ and $\mmseC= [\bI_{MN}\frac{\sigma_{\nu^2}}{\sigma_{d^2}}+ (\bH \bA)^{\rm H} (\bH \bA)]^{-1} (\bH \bA)^{\rm H}$. First term in above equation holds desired plus interference values and second term holds the post processing noise values.
\subsection{SINR Computation}\label{subsection:sinr}
Suppose we want to detect $l^{\rm th}$ symbol. Estimated $l^{\rm th}$ symbol can be given as,
\be \label{eqn:mmse:estimated:symbol}
\hat{d}_{ l} = [\mmseB]_{ l,l} d_{l} + \sum_{r=0, r\neq l}^{MN-1}{[\mmseB]_{l,r} d_{r}} + \sum_{r=0}^{MN-1}{[\mmseC]_{l,r} \nu_{ r}},
\ee
where first term is desired term, second term is interference term and third term is post processed noise term. Using above equation, $E[{\hat{\vert d_l\vert^2}}]= \sigma_d^2[\mmseB]_{ l,l}^2 + \sigma_d^2 \vert \sum_{r=0, r\neq l}^{MN-1}{[\mmseB]_{l,r}}\vert^2 +\sigma_\nu^2 \vert \sum_{r=0}^{MN-1}{[\mmseC]_{l,r}} \vert^2$, where first term is average signal power $P_{Sig,l}$, second term is average interference power $P_{Sig+Inr,l}$ and third term is average post processing noise power $P_{Npp,l}$. Using (\ref{eqn:mmse:rx}), $E[\hat{\dd} \hat{\dd}^{\rm H}]= \sigma_d^2 \mmseB \mmseB^{\rm H}+ \sigma_\nu^2 \mmseC \mmseC^H$, where diagonal values of first matrix term holds average signal plus interference power $P_{Sig+Inr,l}$, diagonal values of second matrix term holds average post processing noise power, $\mmseB \mmseB^{\rm H} =[[\bI_{MN}\frac{\sigma_{\nu^2}}{\sigma_{d^2}}+ (\bH \bA)^{\rm H} (\bH \bA)]^{-1} (\bH \bA)^{\rm H} (\bH \bA)]^2 = \mmseB^2$ and $\mmseC \mmseC^{\rm H}= [\bI_{MN}\frac{\sigma_{\nu^2}}{\sigma_{d^2}}+ (\bH \bA)^{\rm H} (\bH \bA)]^{-1} (\bH \bA)^{\rm H}(\bH \bA)[\bI_{MN}\frac{\sigma_{\nu^2}}{\sigma_{d^2}}+ (\bH \bA)^{\rm H} (\bH \bA)]^{-1}$ . Using this, average signal power $P_{Sig,l}$ and  average interference power$P_{Inr,l}$, average signal plus interference power $P_{Sig+Inr,l}$ and post processed noise power $P_{Npp,l}$ for $l^{\rm th}$ symbol, can be given as,
\begin{eqnarray} \label{eqn:mmse:power:interference}
P_{Sig,l} = \sigma_{d^2} \vert [\mmseB]_{l,l} \vert^2,~P_{Sig+Inr,l} = \sigma_{d^2} \vert [\mmseB \mmseB^{\rm H}]_{l,l} \vert,~P_{Inr,l} = P_{Sig+Inr,l} - P_{Sig,l} ~\text{and}~P_{Npp,l} = \sigma_{\nu^2} \vert [\mmseC \mmseC^{\rm H}]_{l,l} \vert.
\end{eqnarray}
SINR for $l^{\rm th}$ symbol can be computed as,
\be \label{eqn:sinr:l} 
\gamma_l = \frac{P_{Sig,l}}{P_{Inr,l}+P_{Npp,l}}
\ee
Now we will compute SINR for FSFC and AWGN channel separately.
\subsubsection{FSFC}\label{section:ber_mmse:sinr:fsfc}
Both $\mmseB$ and $\mmseC \mmseC^{\rm H}$ involve $(\bH \bA)^{\rm H} (\bH \bA)$. To compute (\ref{eqn:sinr:l}) we explore the product. $\HAHA$ is a Hermitian matrix and hence it can be diagonalized as, $\HAHA = \bV \BLambda \bV^{\rm H}$, where, $\bV$ is a unitary matrix which holds eigenvectors of $\HAHA$ in it's columns and $\BLambda = diag\{\lambda_0,\lambda_1,\cdots \lambda_{\rm MN-1}\}$ is a diagonal matrix which holds eigenvalues of $\HAHA$.
We can write, $\mmseB = \bV \tilde{\BLambda} \bV^{\rm H}$, where, $\tilde{\BLambda}= diag\{\tilde{\lambda}_0,\tilde{\lambda}_1, \cdots \tilde{\lambda}_{\rm MN-1}\}$, $\tilde{\lambda}_s=\frac{\lambda_s}{\snri+\lambda_s}$ and $\mmseC \mmseC^{\rm H} = \bV \bar{\BLambda} \bV^{\rm H}$, where,  $\tilde{\tilde{\BLambda}}= diag\{\tilde{\tilde{\lambda}}_0,\tilde{\tilde{\lambda}}_1, \cdots \tilde{{\lambda}}_{\rm MN-1}\}$ and $\tilde{\tilde{\lambda}}_s=\frac{\lambda_s}{(\snri+\lambda_s)^2}$. Putting values of $\mmseB$  and $\mmseC$ into (\ref{eqn:mmse:power:interference}), we can get,
\begin{eqnarray} \label{eqn:mmse:fsfc:powers:final}
P_{Sig,l} = \sigma_{d^2} \vert [\bV \tilde{\BLambda} \bV^{\rm H}]_{l,l} \vert^2,~P_{Inr,l} = \sigma_{d^2} \vert \sum_{r=0,r\neq l}^{MN-1}{[\bV \tilde{\BLambda} \bV^{\rm H}]_{l,r}} \vert^2 ~\text{and}~P_{Npp,l} = \sigma_{\nu^2} \vert [\bV \tilde{\tilde{\BLambda}} \bV^{\rm H}]_{l,l} \vert^2.
\end{eqnarray}
Through the reduction, complex expression involving matrix inverse as in (\ref{eqn:mmse:estimated:symbol}), is brought to a simpler form (using eigenvalue decomposition of $\HAHA$) i.e. instead of computing $\mmseB$ we can proceed directly with $\bH \bA$. 
\subsubsection{AWGN} \label{section:ber_mmse:sinr:awgn}
In case of AWGN, $\bH=\bI_{MN}$, hence we can write the following,
\begin{eqnarray} \label{eqn:awgn:BC}
\mmseB = [\mI+\AHA]^{-1} \AHA, ~\mmseC = [\mI+\AHA]^{-1} \bA^{\rm H} ~\text{and}~\mmseC \mmseC^{\rm H} = [\mI+\AHA]^{-1} \AHA [\mI+\AHA]^{-1}.
\end{eqnarray}
It can be seen that $\AHA$ is a major component of the analysis. It is important to study the $\AHA$ before we proceed further. We will see the properties of $\ANHA$ below.\\
\\ {\bf{$\ANHA$ }} \\
Matrix, $\bA_N$, can be decomposed as
\be
\bA_N =  \begin{bmatrix}
[\BG_0 \bE]_{MN \times N} & \cdots & [\BG_{M-1} \bE]_{MN \times N}
\end{bmatrix}_{MN \times MN}, 
\ee
 where, $\BG_p$'s be a set of $MN \times MN$ matrices, where $p=0, \cdots , M-1$.  Suppose the first matrix in the set $\BG_0 = diag\lbrace \textit {\bg}^T \rbrace$ , where $\bg= [g(0)~g(1)\cdots g(MN-1)]^{\rm T}$. Any matrix in the set can be written as a circularly shifted version of $\BG_0$ along it's diagonals i.e. $p^{th}$ matrix in the set can be written as $\BG_p = \mathrm{diag}\{\mathrm{circshift}[\bg^T, -pN]\}$, where $\mathrm{circshift}$ represents right circular shift operation. $\BG_p$ can be described as,

\be
 \BG_p = \begin{bmatrix}
 g{(-pN)_{MN}} & & \\
  & \ddots & \\
  & & g{(-pN+MN-1)_{MN}}
 \end{bmatrix}_{MN \times MN}
\ee
and, $\bE= [\bold{\mathbb{W}}_N  \cdots\textsl{M times} \cdots \bold{\mathbb{W}}_N ]^T$ is a $MN \times N $ matrix, where $\bold{\mathbb{W}}_N$ is $N \times N$ normalized inverse DFT matrix\cite{strang}. 
$[\ANHA]_{r,q}= \bE^H \BG_r^H \BG_q \bE$, where $r,q= 0,\cdots, M-1$ and $\bE^H \BG_r^H \BG_q \bE$ is $N \times N$ matrix. $[\ANHA]_{r,q}$ can be written as,
\be \label{L}
[\ANHA]_{r,q}(\alpha,\beta) =\begin{cases} \sum_{\kappa=0}^{MN-1}{b_{\kappa}} & \alpha=\beta\\ \sum_{\kappa=0}^{MN-1}{\omega^{(\beta-\alpha)\kappa}} b_{\kappa} & \text{otherwise} \end{cases},  
\ee
where $\omega = e^{\frac{j 2 \pi }{N}}$ , $b_{\kappa} = g((-rN+\kappa)_{MN}))g((-qN+\kappa)_{MN}))$, $r,q= 0 \cdots M-1$ and $\alpha,\beta = 0 \cdots N-1$.
Let, $s= 0 \cdots N-1$
\be 
\begin{aligned}
{[\ANHA]}_{r,q}((\alpha-s)_{N},(\beta-s)_{N}) &= \sum_{\kappa=0}^{MN-1}{\omega^{(\beta-s-\alpha+s)\kappa}} b_{\kappa}
&= [\ANHA]_{r,q}(\alpha,\beta)
\end{aligned}
\ee
This proves that each block $[\ANHA]_{r,q}$ of $\ANHA$ is circulant. Let, $\varsigma=0, \cdots, M-1$.
\be 
{[\ANHA]}_{(r-\varsigma)_M,(q-\varsigma)_M}(\alpha,\beta) = \sum_{\kappa=0}^{MN-1}{\omega^{(\beta-\alpha)\kappa}} g((-rN+\varsigma N+\kappa )_{MN})g((-qN+\varsigma N+\kappa)_{MN}).
\ee
\normalsize
As $\omega$ is periodic with $N$ and  $g(\beta)$ is periodic with $MN$, ${[\ANHA]}_{(r-\varsigma)_M,(q-\varsigma)_M}(\alpha,\beta) = {[\ANHA]}_{r,q}(\alpha,\beta)$, hence, $\ANHA$ is block circulant matrix with circulant blocks(BCCB) with blocks of size $N \times N$. In the same way it can be proved that $\bA_M^{\rm H}\bA_M$ is also BCCB with blocks of size $M \times M$.\\
Using the properties of BCCB matrix one gets that\cite{trapp1973} (i) Inverse of BCCB matrix is a BCCB matrix, (ii) addition of a diagonal matrix with equal elements to a BCCB matrix is a BCCB matrix, (iii) multiplication of two BCCB matrix is BCCB matrix. Hence, it can be easily proved that $\mmseB$ and $\mmseC \mmseC^{\rm H}$ is also a BCCB in case of AWGN channel.\\
For any BCCB matrix \cite{trapp1973} it can be shown that (i) diagonal elements are identical , (ii) all rows have equal power and (iii) all columns have equal power. Using this it can be then concluded that, $P_{Sig,l} = P_{Sig}$, $P_{Sig+Inr,l} = P_{Sig+Inr}$, $P_{Inr,l} = P_{Inr}$ and $P_{Npp,l}= P_{Npp}$, $\forall l$. Therefore we can write, $[\mmseB]_{l,l} = \frac{trace\{\mmseB\}}{MN}$, $[\mmseB^2]_{l,l} = \frac{trace\{\mmseB^2\}}{MN}$ and $[\mmseC \mmseC^{\rm H}]_{l,l} = \frac{trace\{\mmseC \mmseC^{\rm H\}}}{MN}$. Using this and (\ref{eqn:mmse:power:interference}), we can write,
\begin{eqnarray}
P_{Sig} = \frac{\sigma_d^2}{(MN)^2}trace\{\mmseB\}^2, ~ P_{Sig+Inr} = \frac{\sigma_d^2}{MN}trace\{\mmseB^2\}~\text{and} ~ P_{Npp} = \frac{\sigma_{\nu}^2}{MN} trace\{\mmseC \mmseC^H\}.
\end{eqnarray}
Now, using (\ref{eqn:awgn:BC}) in above equation,
\begin{eqnarray}\label{eqn:mmse:awgn:power:final}
P_{Sig} = \frac{\sigma_d^2}{(MN)^2} \vert \sum_{s=0}^{MN-1}{\frac{\lambda_s}{\lambda_s + \snri}}\vert^2, ~ P_{Sig+Inr} = \frac{\sigma_d^2}{MN} \sum_{s=0}^{MN-1}{\vert \frac{\lambda_s}{\lambda_s + \snri}\vert^2} ~\text{and}~ P_{Npp} = \frac{\sigma_{\nu}^2}{MN}  \sum_{s=0}^{MN-1}{\frac{\lambda_s}{(\lambda_s + \snri)^2}}.
\end{eqnarray}
SINR can be computed as,
\be \label{eqn:mmse:awgn:sinr}
\gamma = \frac{P_{sig}}{P_{Inr+P_{Npp}}}.
\ee
Hence, SINR can be computed using the eigenvalues of $\AHA$. Through the above, because of the BCCB property, we can compute SINR easily than by using the direct form. Inversion of matrix $\mI+\AHA$ needs complexity of $O(M^3N^3)$ whereas eigenvalue computation of $\AHA$ needs complexity of $O(NM\log_2N)$\cite{trapp1973}. Hence, SINR computation using above method is much simpler than using direct matrix computation.
\subsection{BER Computation}
\subsubsection{FSFC Channel}
\begin{figure} [h!]
\begin{subfigure}[b]{0.49\textwidth}
\includegraphics[width = \linewidth]{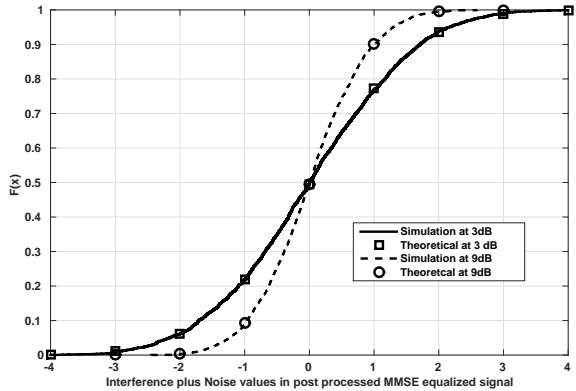}
\caption{FSFC channel}
\label{fig:gfdm:mmse:I+N_CDF_FSFC_QPSK}
\end{subfigure}
~
\begin{subfigure}[b]{0.49\textwidth}
\includegraphics[width = \linewidth]{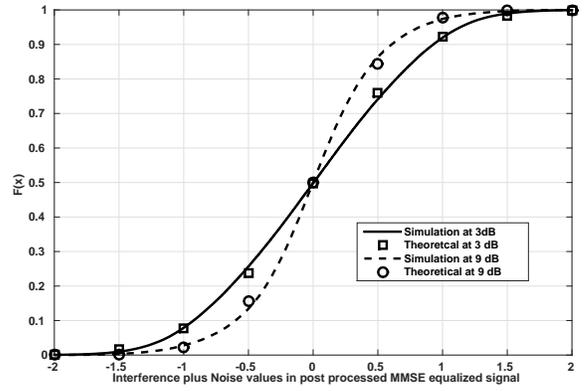}
\caption{AWGN channel}
\label{fig:gfdm:mmse:I+N_CDF_AWGN_QPSK}	
\end{subfigure}
\caption{ CDF plot of interference plus noise value in AWGN Channel and FSFC for 4 QAM (Pulse shaping filter is RRC with ROF =0.5)}
\label{fig:cdf:i+n}
\end{figure}

Fig.~\ref{fig:gfdm:mmse:I+N_CDF_FSFC_QPSK} shows cumulative distribution plot of signal plus interference value for 4 QAM modulated system for 3 dB and 9 dB $\frac{E_b}{N_0}$ values. CDF plot for both cases is compared with Gaussian CDF with measured mean and variance values. It is clear from the figure that interference plus noise values closely follow Gaussian distribution.
Therefore the BER for $l^{th}$ QAM symbol with $\mathcal{M}$ modulation order can be computed as~\cite{simon_digital_2005} by
\begin{align}
 P_b(E|\gamma_l) \simeq 4 & \frac{\sqrt{\mathcal{M}}-1}{\sqrt{\mathcal{M}}log_2(\mathcal{M})}\sum_{r=0}^{\sqrt{\mathcal{M}}/2 -1}\left[Q \lbrace(2r+1)\sqrt{\frac{\gamma_l}{(\mathcal{M}-1)}}\rbrace\right],
\end{align}
$\gamma_l$ is the post processing SINR for $l^{th}$ symbol at the receiver given in (\ref{eqn:sinr:l}). Average probability of error can be found as 
\be \label{eqn:mmse:ber:fsfc}
P_b(E) = \frac{1}{MN} \times \sum_{l=0}^{MN-1}{\int_{0}^{\infty}{P_b(E|\gamma_l) f_{\gamma_l}(\gamma_l)}d\gamma_l},	 
\ee
where, $f_{\gamma_l}(\gamma_l)$ is probability density function of $\gamma_l$.

\subsubsection{AWGN Channel}
Fig.~\ref{fig:gfdm:mmse:I+N_CDF_AWGN_QPSK} shows cumulative distribution plot of interference plus noise value (\ref{eqn:mmse:estimated:symbol}) for 4 QAM modulated system for 3 dB and 9 dB $\frac{E_b}{N_0}$ values. As in case of FSFC, interference plus noise values closely follow Gaussian distribution. BER for QAM symbol with $\mathcal{M}$ modulation order can be computed as \cite{simon_digital_2005},

\begin{align} \label{eqn:mmse:ber:awgn}
 P_b(E) \simeq 4 & \frac{\sqrt{\mathcal{M}}-1}{\sqrt{\mathcal{M}}log_2(\mathcal{M})}\sum_{r=0}^{\sqrt{\mathcal{M}}/2 -1}\left[Q \lbrace(2r+1)\sqrt{\frac{\gamma}{(\mathcal{M}-1)}}\rbrace\right],
\end{align}
where $\gamma$ is the post processing SINR given in (\ref{eqn:mmse:awgn:sinr}).


\section{Precoded GFDM System} \label{sec:gfdm:ietcomm:precoding:precoding}
In this section, precoding schemes to enhance performance of GFDM are described. The first precoding scheme BIDFT is described in Section~\ref{sec:bidft}. The scheme is developed using special properties of $\HAHA$ (detailed in Section~\ref{sec:HAHA}) and $\AHA$ (detailed in Section~\ref{section:ber_mmse:sinr:fsfc}). For this precoding scheme there can be two kinds of receiver processing namely (i) Joint Processing : described in Section~\ref{sec:bidft:joint_processing}, which equalizes channel and GFDM modulation matrix simultaneously, whereas (ii) Two Stage Processing : described in Section~\ref{sec:bidft:two_stage_processing}, first stage equalizes for the channel and second stage equalizes for GFDM modulation matrix. Two types of precoding matrices are defined for BIDFT precoding namely (i) BIDFT-M : where $\bA$ is structured in blocks of $M\times M$ i.e. $\bA_M$ and (ii) BIDFT-N : where $\bA$ is structured in blocks of $N\times N$ i.e. $\bA_N$. BIDFT-N precoded GFDM is processed using Joint Processing as well as Two-Stage Processing whereas BIDFT-M precoded GFDM is processed using Two-Stage Processing only.\\
\indent DFT-based precoding is described in Section~\ref{sec:dft_precoding}. SVD based precoding is described in Section~\ref{sec:svd_precoding}. For each precoding scheme, Precoder matrix $\bP$, corresponding receivers and post processing SNR is described in detail. Both BIDFT and DFT-based precoding does not require channel state information (CSI) at the transmitter to compute $\bP$ whereas SVD based precoding needs CSI at the transmitter to compute $\bP$. Channel knowledge at the transmitter can be maintained via feedback from the receiver or through the reciprocity principle in a duplex system\cite{chung_degrees_2001}. BER for precoded GFDM system is presented in Section~\ref{wpc'14:gfdm:result:precoding}. Computational complexity of GFDM and precoded GFDM system is given in Section~\ref{sec:precoding:computation_complexity}.
\subsection{Block IDFT Precoded GFDM}\label{sec:bidft}
 Received signal in (\ref{eqn: rec}) can be processed in two ways, (i) Joint Processing : Channel and self interference are equalized simultaneously, and, (ii) Two stage processing: Channel and self-interference and equalized separately.
 \subsubsection{Joint Processing}
Suppose the received signal in (\ref{eqn: rec}) passed through a matched filter. Equalized vector which can be given as,
\be {\label{eqn:BDFT:MF}}
\begin{aligned}
\by^{\rm MF} &= (\bH \bA \bP)^{\rm H} \by\\
&= \bP^{\rm H} \HAHA \bP \bd + (\bH \bA)^{\rm H} \bnu.
\end{aligned}
\ee
Since, $\HAHA$ is multiplied to desired data vector in above equation. Now we will explore some properties of $\HAHA$\\
\subsubsection*{$\HAHA$} \label{sec:HAHA}
Using, description of $\bA_N$ given in section~\ref{section:ber_mmse:sinr:fsfc}, $(\bH \bA_N)^H \bH \bA_N$ matrix can be given as,
\be \label{aha_matrix}
(\bH \bA_N)^H \bH \bA_N= \begin{bmatrix}
[{\bold{L}_{0,0}}]_{\rm{N \times N}}& \cdots & [\bold{L}_{0,M-1}]_{N \times N} \\
\vdots & \ddots & \vdots\\
[\bold{L}_{M-1,0}]_{N \times N} & \cdots & [\bold{L}_{M-1,M-1}]_{N \times N} \\
\end{bmatrix}_{MN \times MN},
\ee
where $\bold{L}_{u,v} = \bE^H \BG^H_u \bold{\Upsilon} \BG_{v} \bE$ is a $N \times N$ sub-matrix or block, where $\bold{\Upsilon}= \bH^H \bH$ is a $MN \times MN$ matrix and $u,v= 0 \cdots M-1$. Channel convolution matrix $\bH$ is a circulant matrix and it can be shown that $\bold{\Upsilon}= \bH^H \bH$ is also a circulant matrix using the properties of circulant matrices\cite{strang}. $(\bH \bA_N)^H \bH \bA_N$ will be a block circulant matrix, iff, $\bold{L}_{(u+s)_M,(v+s)_M} = \bold{L}_{u,v}$, where $s= 0 \cdots M-1$. In the expression of $\bold{L}_{u,v}$, matrix $\bE^H$ and $\bE$ are independent of block indices $u,v$. Therefore it can be said that, $\mathbf{G}$ is a block circulant matrix, iff, $\bold{\Phi}_{(u+s)_M,(v+s)_M} = \bold{\Phi}_{u,v}$, where, $\bold{\Phi}_{u,v} = \BG^H_u \bold{\Upsilon} \BG_v $ is $MN \times MN$ matrix. Let, $\bold{\Upsilon} = \{\upsilon_{r,q}\}_{MN \times MN}$ and using the definition of $\BG_p$, it can be shown that
\be
\bold{\Phi}_{u,v} (r,q) = g_{(-uN+r)_{MN} } g_{(-vN+q)_{MN}} \upsilon_{r,q} ~ \text{and}
\ee
\be
\bold{\Phi}_{(u+s)_M,(v+s)_M} (r,q) = g_{(-(u+s)N+r)_{MN} } g_{(-(v+s)N+q)_{MN}} \upsilon_{r,q}.
\ee
Substituting, $r'=r-sN$ and $q' = q-sN$ and since, $\bold{\Upsilon}$ is a circulant matrix and hence $\upsilon_{(r'+sN,q'+sN)_{MN}} = \upsilon_{r',q'} $. Then, substituting $r'$\& $q'$ with $r$ \& $q$,
\begin{eqnarray}
\bold{\Phi}_{(u+s)_M,(v+s)_M} (r,q) = g_{(-uN+r)_{MN} } g_{(-vN+q)_{MN}} \upsilon_{r, q}
=\bold{\Phi}_{(u)_M,(v)_M} (r,q)
\end{eqnarray}
Hence $(\bH \bA_N)^H \bH \bA_N$ is a block circulant matrix with blocks of size $N \times N$. (It may be noted here that $(\bH \bA_M)^H \bH \bA_M$ will not be block circulant with blocks of size $M \times M$). Since $(\bH \bA_N)^H \bH \bA_N$ is block circulant matrix with blocks of size $N \times N$, it can be decomposed as given in \cite{trapp1973,gerlic1983}, as, $(\bH \bA_N)^H \bH \bA_N = \bF_{bN} \bD_{bN} \bF_{bN}^{\rm H}$, where, $\bF_{bN} = \begin{bmatrix}
\bW_N^0 & \bW_N & \cdots \bW_N^{M-1}
\end{bmatrix}_{MN \times MN}$, where, $\bW_N^i =\frac {\begin{bmatrix}
\bI_N &
w_N^i \bI_N &
\cdots &
w_N^{i(N-1)} \bI_N
\end{bmatrix}_{N \times MN}^T} {\sqrt{N}}$,
where,$w_N=e^\frac{j2\pi}{N}$
and $\bD_{bN}= diag\{\bD_{bN}^0~\bD_{bN}^1 \cdots \bD_{bN}^{M-1}\}$ is block diagonal matrix with blocks of size $N\times N$, where $\bD_{bN}^r$ is $r^{th}$ diagonal matrix of size $N\times N$.\\
Using this decomposition and taking $\bA$ as $\bA_N$, matched filter output in (\ref{eqn:BDFT:MF}) can be written as,
\be
\by^{MF} = \bP^{\rm H} \bF_{bN} \bD_{bN} \bF_{bN}^{\rm H} \bP \bd + (\bH \bA_N)^{\rm H} \bnu.
\ee
\subsubsection*{BIDFT-N Precoding} \label{sec:bidft:joint_processing}
if we choose, $\bP = \bF_{bN}$ (call it block inverse discrete Fourier transform -N (BIDFTN) precoding), then,
\be
\begin{aligned}
\by^{MF} &= \bD_{bN} \bd + (\bH \bA_N)^{\rm H} \bnu\\
& = \begin{bmatrix}
\bD_{bN}^0 & & & \\
& \bD_{bN}^1 & & \\
& & \ddots & & \\
& & & \bD_{bN}^{N-1}
\end{bmatrix} \begin{bmatrix}
\bd_0^N \\
\bd_1^N \\
\vdots \\
\bd_{M-1}^N
\end{bmatrix} + \bar{\bnu},
\end{aligned}
\ee
where $\bar{\bnu}$ is MF processed noise vector. In above equation $\bD_{bN}$ being block diagonal matrix, adds only $N-1$ interfering symbols instead $MN-1$ (in case of uncoded GFDM). This shows that precoding reduces the number of interfering symbols significantly. Zero forcing equalization is applied to reduce interference further. Multiplying $\bD_{bN}^{-1}$ in above equation we get,
\be \label{eqn:est:bidftn:jp}
\hat{\bd}^{JP}_{bidft} = \bd + \bnu^{JP}_{bdft},
\ee
where, $\bnu^{JP}_{bdft}= \bD_{bN}^{-1} (\bH \bA_N)^{\rm H} \bnu$ is post processing noise vector and superscript $\textsc{JP}$ signifies that signal processing steps followed in this method are joint processing (channel and self interference are equalized jointly). $\bD_{bN}$ can be computed as,
\be
\bD_{bN} = \bF_{bN}^{\rm H} (\bH \bA_N)^{\rm H} \bH \bA_N \bF_{bN}.
\ee
Post processing SNR for $l^{th}$ symbol can be obtained as,
\be \label{eqn:snr:bdft:jp}
\gamma_{bidft,l}^{JP} = \frac{\sigma_{d^2}}{E[\bnu^{JP}_{bdft} (\bnu^{JP}_{bdft})^{\rm H}]_{l.l}},
\ee
where denominator in above equation is post processing noise power for $l^{th}$ symbol.
\subsubsection{Two Stage Processing} \label{sec:bidft:two_stage_processing}
In above method, channel and GFDM were equalized together. In this method, we will first equalize channel distortions, and then GFDM induced self-interference. As explained in section~\ref{sec:introduction}, $\bH$ is a circulant matrix. Hence, $\bH$ can be decomposed as,
\be
\bH = \bW_{NM} \NFD \bW_{NM}^{\rm H},
\ee
where, $\bW_{NM}$ is normalized IDFT matrix of size $MN \times MN$ and $\NFD = diag\{\upsilon_0,\upsilon_1, \cdots \upsilon_{MN-1}\}$ is a diagonal matrix. Channel equalized vector can be obtained as,
\be
\begin{aligned}
\by_{FDE} &= \bW_{NM}^{\rm H} \NFD^{-1} \bW_{NM} \by &=\bA \bP \bd + \bW_{NM} \NFD^{-1} \bW_{NM}^{\rm H} \bnu,
\end{aligned}
\ee
where first term is the transmitted signal which is free from channel distortions completely, second term is enhanced noise and subscript $FDE$ is acronym for frequency domain equalization (as above described channel equalization is frequency domain equalization \cite{sesia}). $\NFD$ can be obtained as $\NFD = \bW_{NM}^{\rm H} \bH \bW_{NM}$ which equivalently obtained by taking $NM$ point FFT of zero padded channel convolution vector $\bh$ which is also the first column of $\bH$\cite{sesia,holmalte}.\\
Now passing channel equalized data $\by_{FDE}$ to matched filter receiver, we can get,
\be \label{eqn:y:fde:mf}
\begin{aligned}
\by_{FDE-MF} &= (\bA \bP)^{\rm H} \by_{FDE}
&= \bP^{\rm H} \AHA \bP \bd + (\bA \bP)^{\rm H} \bW_{NM} \NFD^{-1} \bW_{NM}^{\rm H} \bnu.
\end{aligned}
\ee
It has been proved in section~\ref{section:ber_mmse:sinr:awgn} that $\AHA$ is BCCB matrix with blocks of size either $N \times N$ or $M \times M$, which depends on whether $\bA = \bA_N$ or $\bA = \bA_M$. Which are described next.
\subsubsection*{BIDFT-N Precoding}
When modulation matrix is defined as $\bA_N$, $\bA_N^{\rm H} \bA_N$ can be decomposed as,
\be
\bA_N^{\rm H} \bA_N = \bF_{bN} \tilde{\bD}_{bN} \bF_{bN}^{\rm H},
\ee
where, $\tilde{\bD_{bN}}= diag\{\bD_{bN}^0,\bD_{bN}^1\cdots \bD_{bN}^{M-1}\}$ is $MN \times MN$ block diagonal matrix where $\bD_{bN}^r$ is $r^{\rm th}$ diagonal block of size $N \times N$. Choosing, $\bP = \bF_{bN}$ (BIDFTN precoding) and using above decomposition, matched filter output in (\ref{eqn:y:fde:mf}) can be written as,
\be
\by_{FDE-MF} = \tilde{\bD}_{bN} \bd + (\bA \bF_{bN})^{\rm H} \bW_{NM} \NFD^{-1} \bW_{NM}^{\rm H} \bnu.
\ee
Now multiplying $\tilde{\bD}_{bN}^{-1}$ in above equation,
\be \label{eqn:est:bidftn:tsp}
\hat{\bd}_{FDE-MF-ZF}^{N} = \bd + \bnu_{FDE-MF-ZF}^N,
\ee
where, $\bnu_{FDE-MF-ZF}^N= \tilde{\bD}_{bN}^{-1} (\bA \bF_{bN})^{\rm H} \bW_{NM} \NFD^{-1} \bW_{NM}^{\rm H} \bnu$ is enhanced noise vector. Since $\tilde{\bD}_{bN}$ needs to be computed to obtain $\hat{\bd}_{FDE-MF_ZF}^{N}$, it can be computed as, $\tilde{\bD}_{bN} = \bF_{bN} \AHA \bF_{bN}^{\rm H}$. Post processing SNR for $l^{th}$ symbol can be obtained as,
\be \label{snr:bdft:N}
\gamma_{FDE-MF-ZF,l}^N = \frac{\sigma_{d^2}}{E[\bnu_{FDE-MF-ZF}^N (\bnu_{FDE-MF-ZF}^N)^{\rm H}]_{l,l}},
\ee
where denominator in above equation is enhanced noise power for $l^{\rm th}$ symbol.
\subsubsection*{BIDFT-M Precoding}
Now, if modulation matrix is defined as $\bA_M$, $\bA_M^{\rm H} \bA_M= \bF_{bM} \bD_{bM} \bF_{bM}^{\rm H} $, where, \\
$\bF_{bM} = \begin{bmatrix}
\bW_M^0 & \bW_M & \cdots \bW_M^{N-1}
\end{bmatrix}_{MN \times MN}$, where, $
\bW_M^r =\frac {\begin{bmatrix}
\bI_M &
w_M^r \bI_M &
\cdots &
w_M^{r(M-1)} \bI_M
\end{bmatrix}_{M \times MN}^T} {\sqrt{M}}$, where,$w_M=e^\frac{j2\pi}{M}$
and $\bD_{bM}= diag\{\bD_{bM}^0~\bD_{bM}^1 \cdots \bD_{bM}^{N-1}\}$ is block diagonal matrix with blocks of size $M\times M$, where $\bD_{bM}^r$ is $r^{th}$ diagonal matrix of size $M\times M$. Using this decomposition, choosing $\bP = \bF_{bM}$ (block IDFT-M (BIDFTM) Precoding) and following same signal processing steps as in case of $\bA_N$, equalized data vector can be given as,
\be \label{eqn:est:bidftm}
\hat{\bd}_{FDE-MF_ZF}^{M} = \bd + \bnu_{FDE-MF-ZF}^M ,
\ee
where, $\bnu_{FDE-MF-ZF}^M =\bD_{bM}^{-1} (\bA \bF_{bM})^{\rm H} \bW_{NM} \NFD^{-1} \bW_{NM}^{\rm H} \bnu$ is enhanced noise vector. Since $\bD_{bM}$ needs to be computed to obtain $\hat{\bd}_{FDE-MF_ZF}^{M}$, it can be computed as,
\be
\bD_{bM} = \bF_{bM} \AHA \bF_{bM}^{\rm H}.
\ee
Post processing SNR for $l^{th}$ symbol can be obtained as,
\be \label{snr:bdft:M}
\gamma_{FDE-MF-ZF,l}^M = \frac{\sigma_{d^2}}{E[\bnu_{FDE-MF-ZF}^M (\bnu_{FDE-MF-ZF}^M)^{\rm H}]_{l,l}},
\ee
where denominator in above equation is enhanced noise power for $l^{\rm th}$ symbol.
\begin{figure}[h]
\centering
\begin{subfigure}[b]{0.7\linewidth}
\psfrag {AA} {$\Scale [0.7]{\bd}$}
\psfrag{BB} [C][B]{$\Scale [0.7]{\bF_{bN}}$}
\psfrag{ZZ}[C][B]{$\Scale [0.7]{\bF_{bM}}$}
\psfrag {CC} [C][B]{$\Scale [0.7]{\tilde{\dd}}$}
\psfrag {DD} [C][B]{$\Scale [0.7]{\bA}$}
\psfrag {EE} [C][B]{$\Scale [0.7]{\bx}$}
\psfrag {FF}[C][B] {$\Scale [0.7]{\bH}$}
\psfrag {GG} {$\Scale [0.7]{\bnu}$}
\psfrag {HH} [C][B] {$\Scale [0.7]{(\bF_b \bH \bA)^{\rm H}}$}
\psfrag {II} [R][B] {$\Scale [0.7]{\bD_b^{-1}}$}
\psfrag {JJ}[L] {$\Scale [0.7]{\hat{\dd}^{JP}_{bdft}}$}
\psfrag {KK} [C][B] {$\Scale [0.7]{\bH^{-1}}$}
\psfrag {LL} [C][B]{\small{$\Scale [0.7]{\by_{\texttt{FDE}}}$}}
\psfrag {MM} [C][B]{$\Scale [0.7]{(\bF_b \bA)^{\rm H}}$}
\psfrag {NN}[C][B] {\small{$\Scale [0.7]{\tilde{\bD_b}^{-1}}$}}
\psfrag {OO} [C][B] {$\Scale [0.7]{\hat{\dd}_{\texttt{FDE-MF-ZF}}}$}
\includegraphics[width=\linewidth]{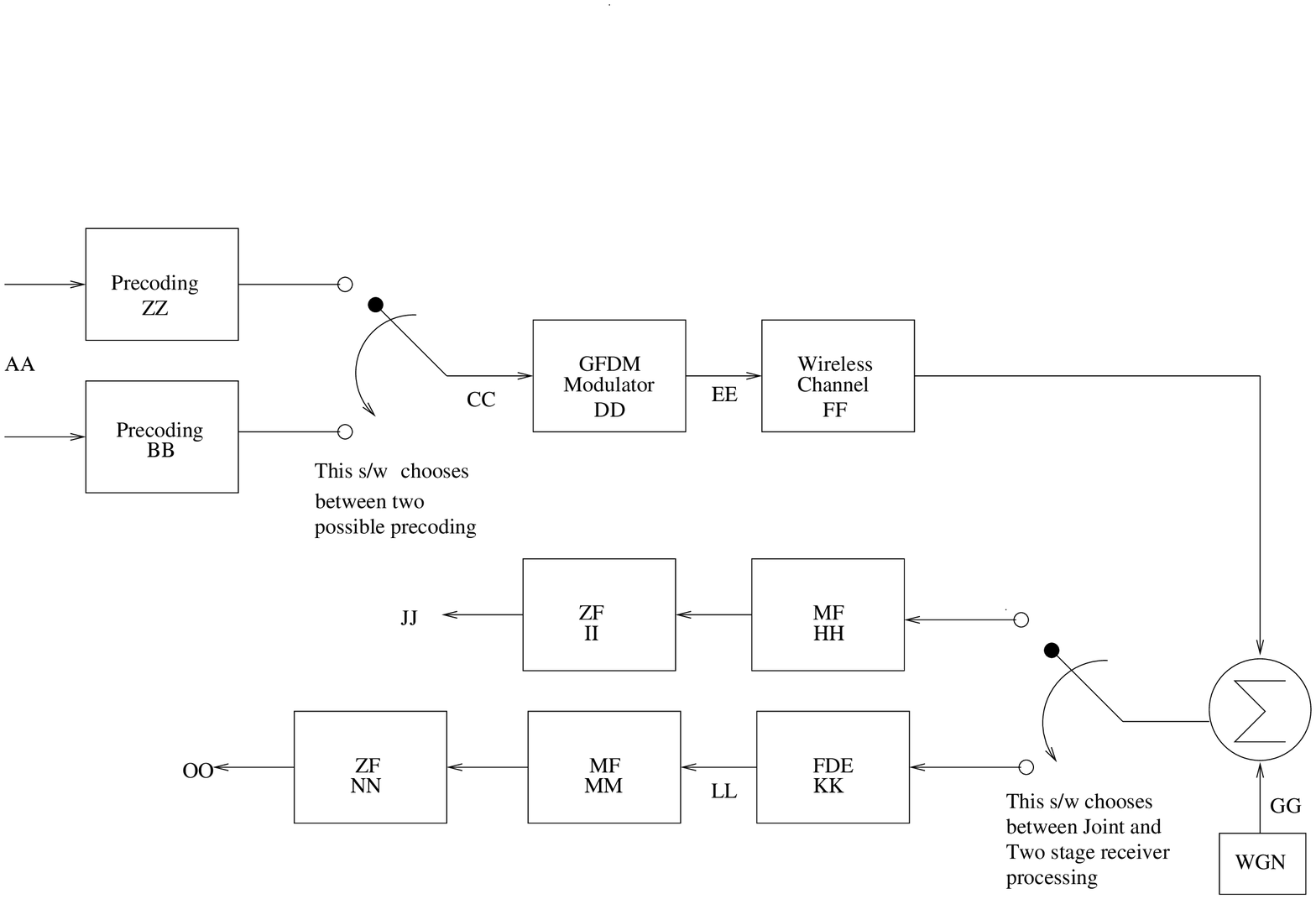}
\caption{BIDFT Precoded GFDM}
\label{fig:bidft}
\end{subfigure}

\begin{subfigure}[b]{0.7\linewidth}
\psfrag{DD}[C][B] {\tiny{ \hspace{0.15cm} $\Scale [0.5]{\bW_{N_{\texttt{DFT}}}}$}}
\psfrag{GG} [C][B]{$\Scale [0.7]{\bP_m}$}
\psfrag{HH} [C][B]{$\Scale [0.7]{\bA}$}
\psfrag {II}[C][B] {$\Scale [0.7]{\bx}$}
\psfrag{JJ} [C][B]{$\Scale [0.7]{\bH}$}
\psfrag {LL}[C][B] {$\Scale [0.7]{\bnu}$}
\psfrag {MM} {}
\psfrag {NN} [C][B]{{$\Scale [0.7]{\bH^{-1}}$}}
\psfrag {OO} [C][B]{$\Scale [0.7]{\by_{\texttt{FDE}}}$}
\psfrag {QQ} [C][B]{$\Scale [0.7]{\bP_m^{\rm H}}$}
\psfrag {Precoding}[C][B] {\hspace{1cm}\small{\Scale [0.7]{\bf{Processing}}}}
\psfrag {PP} {}
\psfrag {RR} [C][B] {\hspace{0.20 cm}\tiny{$\Scale [0.5]{\bW_{N_{\texttt{DFT}}}^{\rm H}}$}}
\psfrag {SS}[C][B] {\hspace{0.25cm}\tiny{$\Scale [0.5]{\bW_{N_{\texttt{DFT}}}^{\rm H}}$}}
\psfrag {TT}[C][B] {\hspace{0.27cm}\tiny{$\Scale [0.5]{\bW_{N_{\texttt{DFT}}}^{\rm H}}$}}
\includegraphics[width = \linewidth]{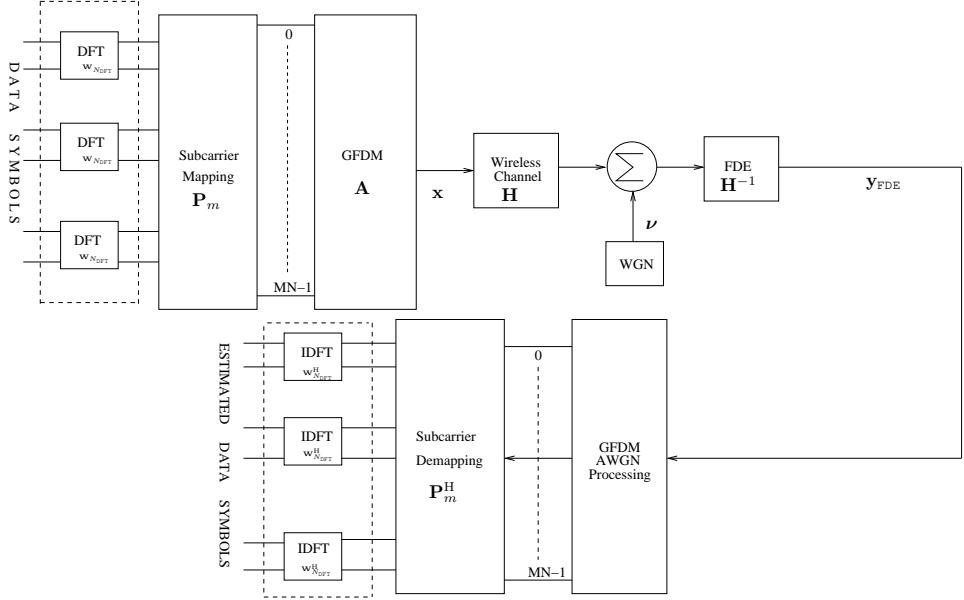}
\caption{DFT Precoded GFDM}
\label{fig:dft_spread_gfdm}
\end{subfigure}
\caption{Block Diagram of BIDFT and DFT precoded GFDM System}
\label{figure:precoded}
\end{figure}
In summary, BIDFT precoding can be understood by Fig.~\ref{fig:bidft}.

\subsection{DFT Precoded GFDM} \label{sec:dft_precoding}
DFT precoding has been used in OFDM systems\cite{sesia,holmalte}. It has been shown that DFT precoding reduces PAPR significantly and is one of the optimum precoding matrices to reduce PAPR in OFDM systems \cite{falconer2011}. This motivates us to investigate DFT precoded GFDM for reducing PAPR in GFDM system. Suppose, $Q$ is spreading factor of the system then DFT order/ size can be computed as, $N_{DFT} = \frac{N}{Q}$. We assume that $Q$ divides $N$ completely. Two subcarrier mapping schemes are considered in this work \cite{myung_peak--average_2006} (i) Localized frequency division multiple access (LFDMA) and (ii) Interleaved frequency division multiple access (IFDMA).\\
Precoding matrix $\bP$ can be defined as $\bP = \bP_m \bP_c$, where $\bP_c$ is a block diagonal matrix with each block being a DFT spreading matrix and $\bP_m$ is a permutation matrix which implements subcarrier mapping i.e. LFDMA or IFDMA. The precoding spreading matrix $\bP_c$ can be written as,
\be
\bP_c = \begin{bmatrix}
\bW_{N_{DFT}} & & & \\
& \bW_{N_{DFT}} & & \\
& & \ddots & \\
& & & \bW_{N_{DFT}}
\end{bmatrix},
\ee
where, $\bW_{N_{DFT}}$ is normalized DFT matrix of size $N_{DFT} \times N_{DFT}$. Permutation matrix, $\bP_m$ for LFDMA is an identity matrix. DFT precoded GFDM system can be understood from Fig.~\ref{fig:dft_spread_gfdm}.
Precoded data vector is GFDM modulated using the modulation matrix $\bA$. Received signal can be equalized using conventional linear\cite{michailow_bit_2012,GFDM_tr} or non-linear\cite{datta_gfdm_2012} equalizer. We will present here ZF receiver for DFT precoded GFDM. ZF equalized precoded data vector can be obtained as,
\be
\begin{aligned}
\hat{\tilde{\bd}}_{zf} &= \HA^{-1} \by
&= \tilde{\bd} + \HA^{-1} \bnu.
\end{aligned}
\ee
Equalized data vector $\hat{\dd}_{dft-spread}$ can be obtained as,
\be
\begin{aligned}\label{eqn:est:dft}
\hat{\dd}_{dft-spread} &= \bP^{\rm H} \hat{\tilde{\dd}}_{zf}
&= \underbrace{\bP_c^{\rm H}}_{De-spreading} \times \underbrace{\bP_m^{\rm H}}_{Subcarrier De-mapping} \times \hat{\tilde{\dd}}_{zf}
&= \bd + \bnu_{dft-zf},
\end{aligned}
\ee
where, $\bnu_{dft-zf}= \bP^{\rm H}(\HA)^{-1} \bnu $ is post processing noise vector. Post processing SNR for $l^{th}$ symbol can be obtained as,
\be \label{snr:dft}
\gamma_{dft-ZF,l}^N = \frac{\sigma_{d^2}}{E[\bnu_{dft-zf}^N (\bnu_{dft-zf}^N)^{\rm H}]_{l,l}},
\ee
where denominator in above equation is enhanced noise power for $l^{\rm th}$ symbol.
\subsection{SVD Precoded GFDM} \label{sec:svd_precoding}
%
The product of the channel matrix and the modulation matrix ($\bH \bA$) can be decomposed as,
\begin{equation}
\bH \bA = \bU \bS \bV^H,
\end{equation}
where $\bU$ and $\bV$ are unitary matrix and $\bS$ is diagonal singular-value matrix i.e. $\bS = diag\{s_0,~s_1\cdots s_r, \cdots s_{MN-1}\}$, where $s_r$ is $r^{th}$ singular value.
Then, $\by$ in (\ref{eqn: rec}), can be written as,
\be
\by = \bU \bS \bV^H \bP \dd + \bnu.
\ee
At transmitter, by choosing $\bP = \bV$ (assuming ideal feedback channel) and multiplying both sides with $\bU^H$, the estimated symbol vector can be written as,
\be
\begin{aligned}
\hat{\dd}^{svd} &= \bS \dd + \bU^H \bnu.
\end{aligned}
\ee
The estimated $l^{th}$ symbol is then given by,
\begin{equation}\label{eqn:est:svd}
\hat{d}_{l}^{svd} = s_l d_l + \sum_{q=0}^{MN-1}{[\bU^H]_{l,q} \nu(q)}.
\end{equation}
From the above, it can be seen that by using SVD based precoding, interference is completely removed by orthogonalization of $\bH \bA$ without the need for matrix inversion, which is required in zero forcing (ZF) and minimum mean square error (MMSE) receiver. SINR for $l^{\mathrm{th}}$ symbol, can be computed as,
\be \label{eqn:snr_svd}
\gamma_{l}^{svd} = \frac{\sigma_d^2} {\sigma_\nu^2}\vert s_l \vert^2.
\ee
\subsection{BER Performance of Precoding Techniques} \label{wpc'14:gfdm:result:precoding}
Expression for estimated data symbols, given in(\ref{eqn:est:bidftn:jp},\ref{eqn:est:bidftn:tsp},\ref{eqn:est:bidftm},\ref{eqn:est:dft},\ref{eqn:est:svd}), are summations of desired data symbols and enhanced noise vector. The enhanced noise vector is weighted sum of complex Gaussian random variable for a given channel realization. Hence, enhanced noise is also complex Gaussian random vector for a given channel realization. BER for QAM symbol with modulation order $\mathcal{M}$ over FSFC can be obtained as,
\begin{align}\label{eqn:ber:precoding}
P_b(E|\gamma_l) \simeq 4 & \frac{\sqrt{\mathcal{M}}-1}{\sqrt{\mathcal{M}}log_2(\mathcal{M})}\sum_{r=0}^{\sqrt{\mathcal{M}}/2 -1}\left[Q \lbrace(2r+1)\sqrt{\frac{\gamma_l}{(\mathcal{M}-1)}}\rbrace\right],
\end{align}
where, $\gamma_l$ is post processing SNR computed in (\ref{eqn:snr:bdft:jp},\ref{snr:bdft:N},\ref{snr:bdft:M},\ref{snr:dft},\ref{eqn:snr_svd}). Average probability of error can be found as
\be
P_b(E) = \frac{1}{MN} \times \sum_{l=0}^{MN-1}{\int_{0}^{\infty}{P_b(E|\gamma_l) f_{\gamma_l}(\gamma_l)}d\gamma_l},
\ee
where $f_{\gamma_l}(\gamma_l)$ is probability distribution function of SINR for $l^{th}$ symbol. 

\subsection{ Computational Complexity}\label{sec:precoding:computation_complexity}
 Precoding schemes for GFDM are proposed in previous subsections. It is utmost important to compute and compare the complexity of these systems as it is directly proportional to the cost of the system. Out of different mathematical operations number of complex multiplication is a significant contributor to computational complexity \cite{blahut10fast}. Computation complexity is computed in terms of number of complex multiplication required to implement transmitter and receiver of precoded and uncoded GFDM system. It has been assumed that modulation matrix $\bA$ is known at the receiver, hence any matrix that is derived from $\bA$ is also known to receiver, such as, $\bA^{\rm H}$, $\bA^{-1}$, etc. All known receivers for uncoded GFDM are considered for complexity computation of uncoded GFDM and DFT precoded GFDM. Minimum number of complex multiplication required to perform various matrix and vector operations is computed . Complexity of SVD precoded GFDM receiver is computed for two cases namely (i)SVD of channel is (useful when channel is static for multiple transmit instances) and (ii) SVD of channel is unknown. Complexity computation can be found following Table~\ref{tab:GFDM:precoding:complexity}.
\begin{table*}
\begin{center}
\footnotesize{
\begin{tabular}{|p{2cm}|p{10cm}|p{3cm}|}
\hline
\bf{Technique} & \bf{Operations} & \bf{Number of Complex Multiplications}\\ \hline
GFDM Tx &
one vector matrix multiplication for $\by = \bA \dd$ & $(MN)^{2}$ \\ \hline 
GFDM Rx ZF-MF & \begin{enumerate} \item Frequency Domain equation (ZF) \begin{itemize} \item one FFT operation for $\by'= \bW \by$. \item one diagonal complex valued matrix inversion and one diagonal matrix and vector multiplication for $\by'' = \BLambda^{-1} \by'$. \item one FFT operation for $\by_{FDE} = \bW \by''$
\end{itemize}  \item Matched Filter for AWGN : one Matrix Vector multiplication for $\hat{d}_{ZF}= \bA^H \by_{FDE}$\end{enumerate} & $\frac{3MN}{2}\log_2(MN)+ 2MN +(MN)^2$  \\ \hline

GFDM Rx ZF-ZF & \begin{enumerate} \item Frequency Domain equation (ZF): same operations as in ZF-MF. \item ZF for AWGN : one Matrix Vector multiplication for $\hat{d}_{MF}= \bA^{-1}\by_{FDE}$. Considering $A^{-1}$ to be known and precomputed at the receiver.\end{enumerate} &$\frac{3MN}{2}\log_2(MN)+ 2MN +(MN)^2$\\ \hline

GFDM Rx ZF-MMSE & \begin{enumerate} \item Frequency Domain equation (ZF): same operations as in ZF-MF. \item MMSE for AWGN \begin{itemize} \item MN complex value addition for computing $\bC = \frac{\bI}{\gamma} + \bA^H \bA$. \item one matrix inversion for computing $\bC^{-1}$. \item one matrix vector multiplication for $\by_{temp} = \bA^H \by$. \item one matrix vector multiplication for $\hat{\dd}_{MMSE} = \bC^{-1} \by_{temp}$. 
\end{itemize} \end{enumerate} &$\frac{3MN}{2}\log_2(MN)+ \frac{(MN)^3}{3}+2(MN)^2 + \frac{2MN}{3}$\\ \hline

GFDM Rx ZF-SIC & \begin{enumerate} \item Frequency Domain equation (ZF): same operations as in ZF-MF. \item DSIC for AWGN \begin{itemize} \item MF operation to compute $\by_{MF}$ and $MN (\sqrt{modorder}-1)$ comparators for detection of matched filter $\hat{\dd}_{detect}$.  \item $2M$ complex multiplication and $2M-1$ complex addition are needed for each sub-carrier and iteration index to compute $\by_{k} = \bA \dd_k$ where $k$ is sub-carrier index. \item $MN$ complex subtraction are needed for each sub-carrier and iteration index to compute $\by_{interfree,k} = \by_{MF}- \by_{k}$  \end{itemize} \end{enumerate} & $\frac{3MN}{2}\log_2(MN)+2MN+2(MN)^2J$\\ \hline

SVD precoded GFDM Tx & two vector matrix multiplication for computing $\bA \bV \dd$ & $2(MN)^2$  \\ \hline

SVD precoded GFDM Rx (Known SVD) &  one vector matrix multiplication for computing $\bU^H \by$ &$(MN)^2$ \\ \hline

SVD precoded GFDM Rx (Un-Known SVD) & \begin{enumerate} \item SVD computation \item one vector matrix multiplication for computing $\bU^H \by$ \end{enumerate} &$(MN)^2+ 26 (MN)^3$ \\ \hline

BIDFT precoded GFDM Tx& one vector matrix multiplication for computing $\bA \bF_b \dd$ as $\bA \bF_b$ can be precomputed at transmitter & $(MN)^2$ \\ \hline

BIDFT precoded GFDM Rx (Joint Processing) & \begin{enumerate} \item one vector matrix multiplication to compute $\bH \bA \bF_b$  \item Computation of block diagonal matrix $\bD_b$. \item Inversion $\bD_b^{-1}$ which can be computed by inverting M square matrices of order $N$.  \end{enumerate} &$(MN)^2\log_2(N)+2(MN)^2+MN^2$ \\ \hline

BIDFT precoded GFDM Rx (ZF-ZF) & \begin{enumerate}\item Frequency Domain Equalization : Same as in ZF-MF.
\item AWGN Processing : \begin{itemize} \item one vector matrix multiplication for computing $\by_{MF} = (\bA \bF_b)^H \by$. \item Computation of block diagonal matrix $\bD_b^{-1}$. \item $N$ times square matrix inversion of order $M$ and block diagonal matrix and vector multiplication to compute $\hat{\dd}_{BDFT} = \bD_b^{-1} \by_{MF}$
\end{itemize} \end{enumerate} & $ \frac{3MN}{2} \log_2(MN)+ 2MN + NM^2 + (MN)^2+ \underbrace{NM^2}_{\rm BIDFTM} \text{or} \underbrace{MN^2}_{\rm BIDFTN}$ \\ \hline

DFT Precoded GFDM Tx (additional over GFDM Tx) & \begin{enumerate}
\item additionally, $MQ$ times $N_{DFT}$ point FFT and Sub-carrier Mapping \item same operation as in GFDM Tx \end{enumerate}& $\frac{MN}{2} \log_2{N_{DFT}}$\\ \hline
DFT Precoded GFDM Rx (additional over GFDM Rx) &\begin{enumerate}
\item same operation as in GFDM Rx.
\item additionally,$MQ$ times $N_{DFT}$ point FFT and Sub-carrier Mapping \item same operation as in GFDM Tx \end{enumerate}& $\frac{MN}{2} \log_2{N_{DFT}}$\\ \hline
\end{tabular} } 

\caption{Number of Complex Multiplication of different techniques in GFDM}
\label{tab:GFDM:precoding:complexity}
\end{center}
\end{table*}


\section{Results}\label{sec:result}
In this section, the results related to works described in earlier sections are presented. Analytical evaluation of BER of MMSE receiver with GFDM is given in Section~\ref{sec:result:ber_mmse}. BER evaluation of precoded GFDM system is provided in Section~\ref{sec:ber:precoding}. Complexity of different transmitters and receivers of GFDM and precoded GFDM is given in Section~\ref{sec:result:complexity}. Finally, PAPR of proposed precoding schemes is compared with GFDM and OFDM in Section~\ref{sec:result:papr}. GFDM system with parameters given Table~\ref{tab:simu:para:mmse} is considered here. It is assumed that the subcarrier bandwidth is larger than the coherence bandwidth of the channel for FSFC. SNR loss due to CP is also considered for FSFC.
\small{\begin{table} [H]
\centering
\begin {tabular}{|c|c|}
\hline
Number of Subcarriers $N$ & 128\\ \hline
Number of Timeslots $M$ & 5 \\ \hline
Mapping & 16 QAM \\ \hline
Pulse shape & RRC with ROF =0.1 or 0.5 or 0.9\\ \hline
CP length $N_{CP}$ & 16 \\ \hline
Channel & AWGN and FSFC \\ \hline
Channel Length $N_{ch}$ & 16 \\ \hline
Power delay profile & $[10^{-\frac{\alpha}{5}}]^{\rm T} $, where $\alpha = 0,1\cdots N_{ch}-1$ \\ \hline
Sub-carrier Bandwidth & 3.9 KHz \\ \hline
RMS delay Spread & 4.3 $\mu$ sec \\ \hline
Coherence Bandwidth & 4.7 KHz \\ \hline
\end{tabular}
\caption{Simulation Parameters of GFDM receivers}
\label{tab:simu:para:mmse}
\end{table}}
\normalsize
\subsection{BER evaluation of GFDM with MMSE receiver}\label{sec:result:ber_mmse}
\begin{figure} [h]
\includegraphics[width = \linewidth]{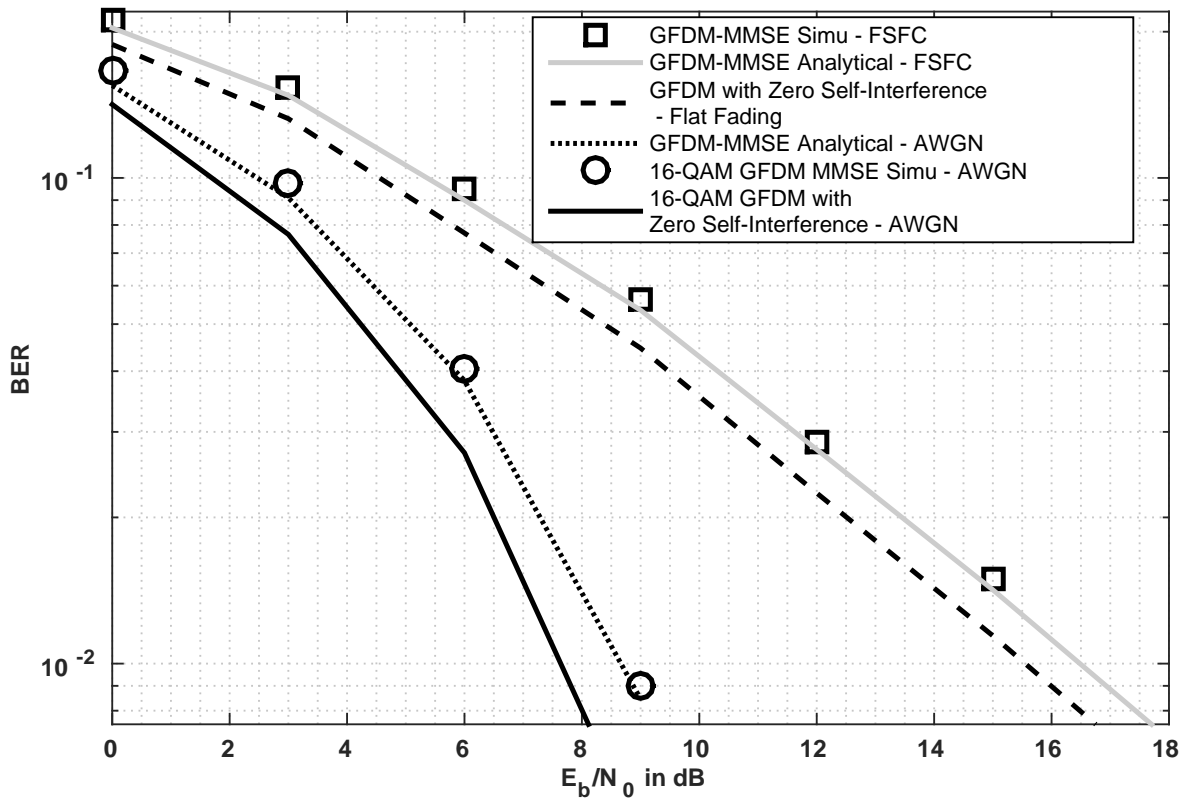}
\caption{ BER vs $\frac{E_b}{N_o}$ for MMSE GFDM Receiver over AWGN and Frequency Selective Channel for 16 QAM (RRC pulse shaping filter is used with ROF=0.5)}
\label{fig:gfdm:mmse:BER_EbNo_FSFC_QPSK_16QAM}
\end{figure}

\noindent BER vs $\frac{E_b}{N_o}$ for 16-QAM in AWGN and FSFC for MMSE receiver is presented in Fig.~\ref{fig:gfdm:mmse:BER_EbNo_FSFC_QPSK_16QAM}. Legends `GFDM without Self-Interference' are obtained by using $\gamma_l=\frac{\sigma_d^2}{\sigma_\nu^2}$ in (\ref{eqn:mmse:ber:fsfc}) for flat fading and in (\ref{eqn:mmse:ber:awgn}) for AWGN channel. This is a reference BER result for  GFDM with zero self-interference and 16 QAM modulation in flat fading (\ref{eqn:mmse:ber:fsfc}) and AWGN (\ref{eqn:mmse:ber:awgn}). The legends marked `Analytical' are obtained by first computing $\gamma_l$ using (\ref{eqn:sinr:l}) for FSFC and using (\ref{eqn:mmse:awgn:sinr}) for AWGN and then using this $\gamma_l$ in (\ref{eqn:mmse:ber:fsfc}) and (\ref{eqn:mmse:ber:awgn}) to obtain average BER for FSFC and AWGN respectively. The legends marked `Simu' are obtained using Monte Carlo simulations using parameters given in Table~\ref{tab:simu:para:mmse}. It is seen that BER from simulation matches quite well with with the analytical for both FSFC and AWGN. The gap between GFDM curve (analytical and simulation) and GFDM with no self interference curve is due to self interference encountered in GFDM \cite{gaspar_low_2013}. From the above discussion, it can be concluded that the expressions developed in this work are useful in estimating the theoretical BER for MMSE based GFDM receiver in FSFC and AWGN channel.
\subsection{BER Evaluation of Precoded Techniques}\label{sec:ber:precoding}
\begin{figure} [h]
\includegraphics[width = \linewidth]{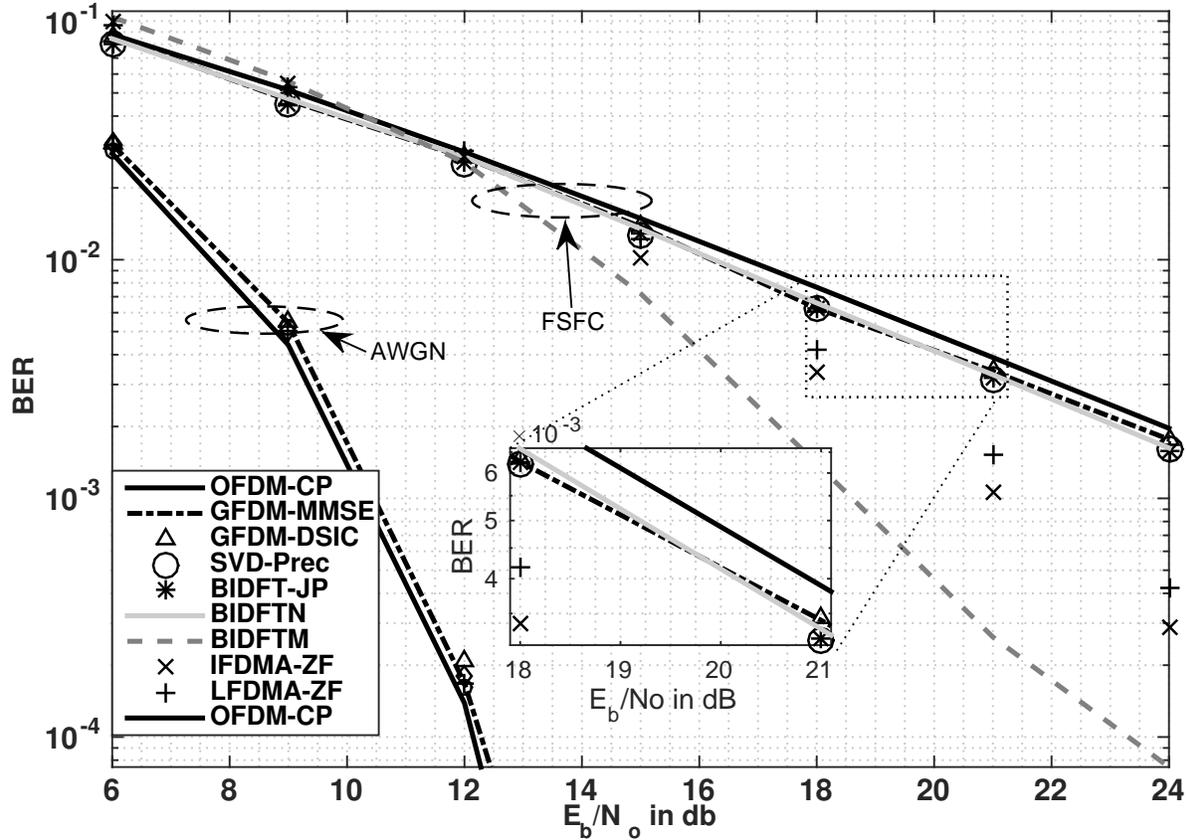}
\caption{ BER vs $\frac{E_b}{N_o}$ for Precoded GFDM Receiver over Frequency Selective and AWGN Channel using 16-QAM with $N$=128,$M$=5, $ROF$=0.1 (RRC)}\label{fig:gfdm:letter:precoding:BER_EbNo_FSFC_16QAM_Precoding_rf01N128M5}
\end{figure}

\begin{figure} [h]
\includegraphics[width = \linewidth]{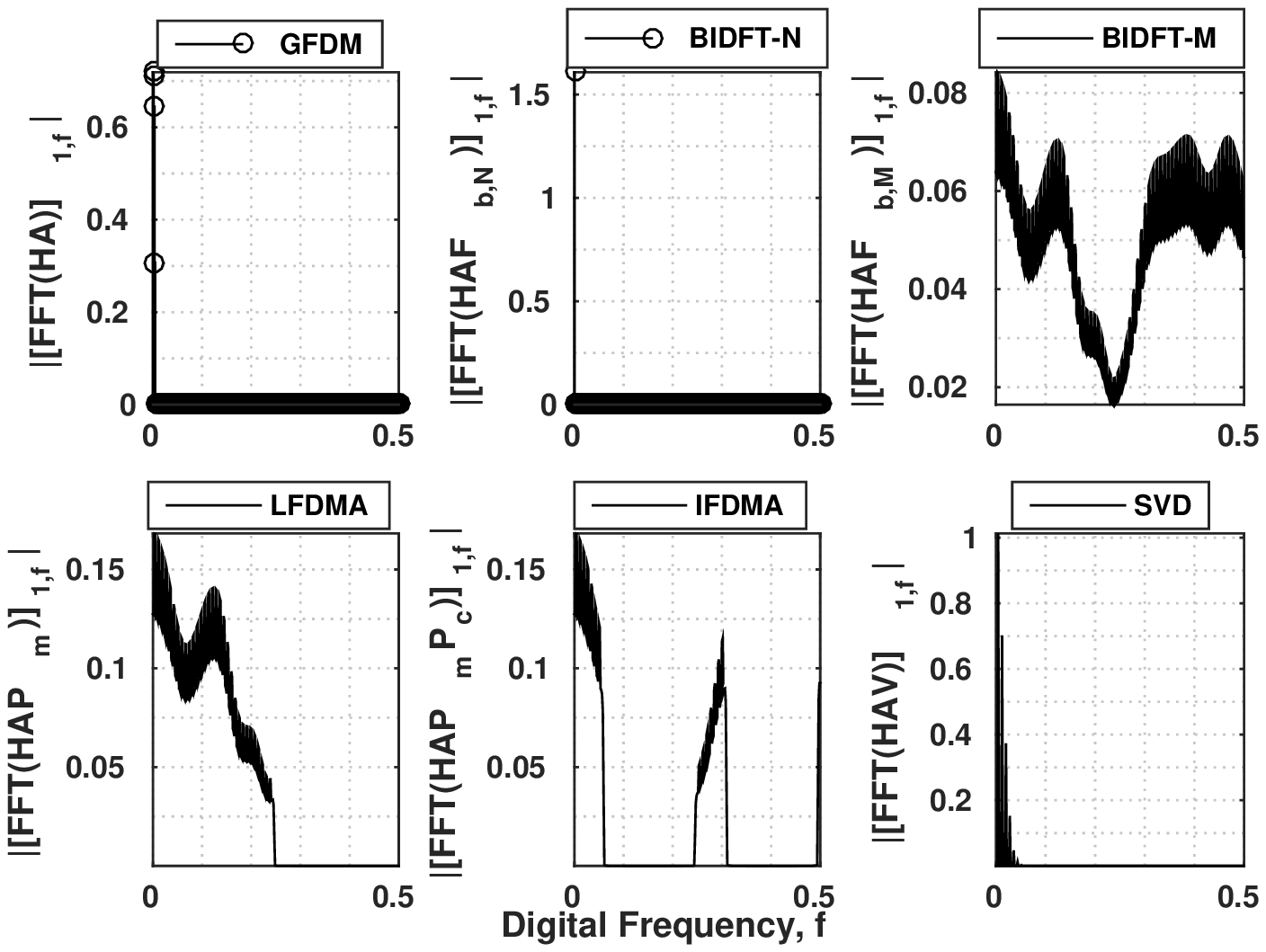}
\caption{ Frequency spread of one symbol for different transmission schemes }\label{fig:frequency_diversity_proof}
\end{figure}

\begin{figure} [h]
\includegraphics[width = \linewidth]{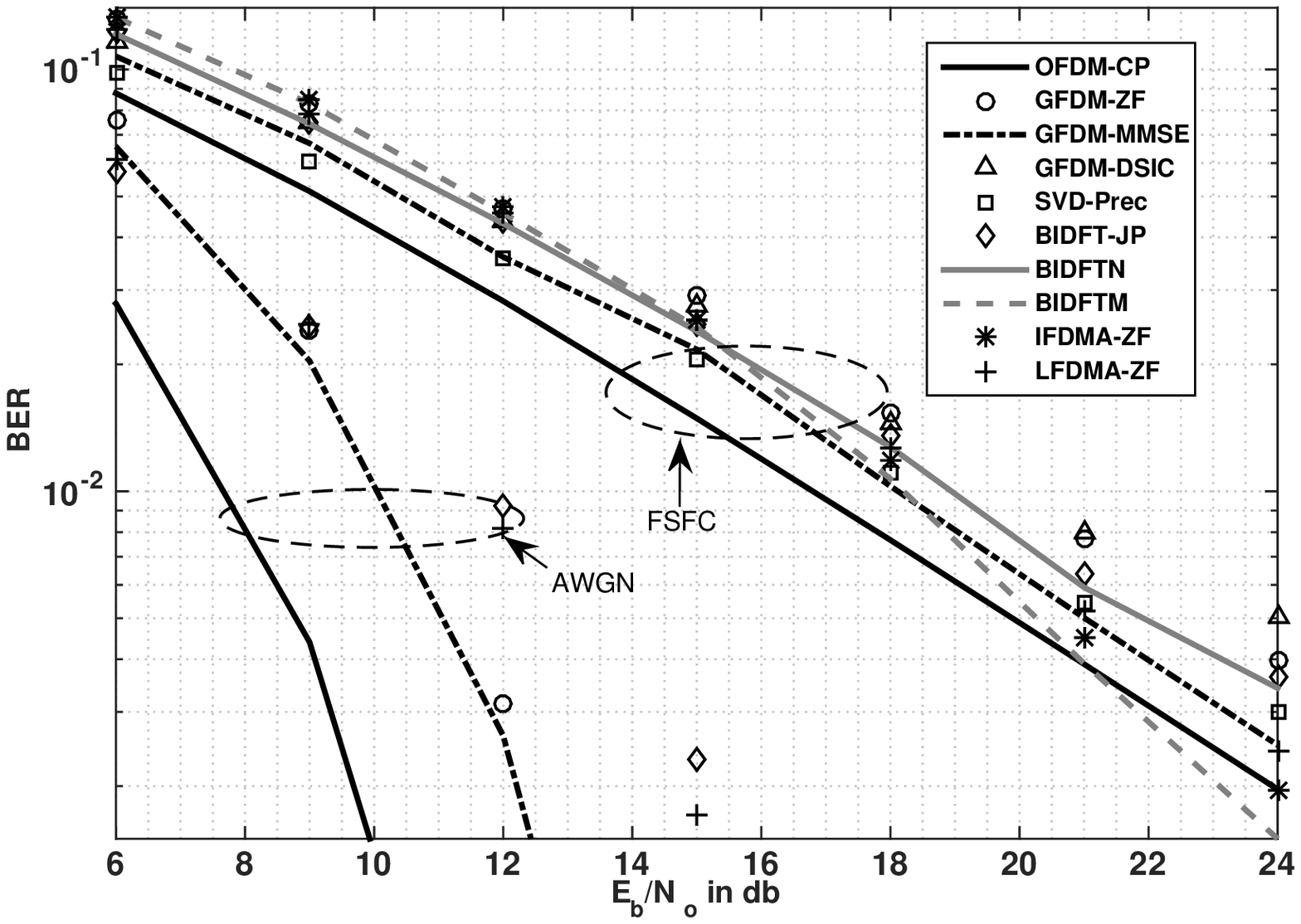}
\caption{ BER vs $\frac{E_b}{N_o}$ for Precoded GFDM Receiver over Frequency Selective and AWGN Channel using 16-QAM with $N$=128,$M$=5, $ROF$=0.9 (RRC)}\label{fig:gfdm:letter:precoding:BER_EbNo_FSFC_16QAM_Precoding_rf09N128M5}
\end{figure}

BER of precoded GFDM system is evaluated via Monte-Carlo simulations. Simulation parameters are given in Table~\ref{tab:simu:para:mmse}. One thousand channel realizations are used to obtain The BER results.
The legends GFDM-ZF, GFDM-MMSE and GFDM-DSIC are used to represent the performance of the corresponding receivers for GFDM. The legends SVD-Prec, BIDFTN, BIDFTM, BIDFT-JP, LFDMA-ZF and IFDMA-ZF are used to indicate the result of SVD based precoding, block IDFTN precoding with two stage processing, block IDFTM with two stage processing, block IDFTN with joint processing, DFT precoding with IFDMA and DFT precoding with LFDMA respectively. Legend OFDM-CP is used to indicate theoretical BER performance of OFDM-CP.
Fig.~\ref{fig:gfdm:letter:precoding:BER_EbNo_FSFC_16QAM_Precoding_rf01N128M5} shows BER vs $\frac{E_b}{N_o}$ for ROF = 0.1 in AWGN channel and FSFC. The following observations can be made.
Under AWGN, all schemes have similar performance. It may be noted that in AWGN, there is no SNR loss as CP is not required. The performance of GFDM is only slightly worse than OFDM. In the case of FSFC, it is seen that SVD, BIDFT-N based precoding has performance similar to OFDM, as ICI is low because of small ROF. The better performance of DFT and BIDFT-M precoded GFDM over OFDM can be attributed to frequency diversity gain, which can be understood from Figure~\ref{fig:frequency_diversity_proof}.
In this figure, it can be seen that one symbol in BIDFT-M has larger frequency spread compared to IFDMA which is larger than LFDMA. Accordingly, BIDFTM precoded GFDM is performing better than IFDMA precoded GFDM which is performing better than LFDMA precoded GFDM.\\
\indent Next we look at Figure~\ref{fig:gfdm:letter:precoding:BER_EbNo_FSFC_16QAM_Precoding_rf09N128M5} which shows similar performance analysis for ROF 0.9, which indicates higher ICI. In case of AWGN, degradation can be observed as compared to Figure~\ref{fig:gfdm:letter:precoding:BER_EbNo_FSFC_16QAM_Precoding_rf01N128M5}. We see a similar degradation for all precoding schemes in FSFC. This is expected for ROF 0.9 as significant overlapping of pulses in frequency.\\
\indent It is clear that the performance of precoding schemes is sensitive to ROF. In this work we have only considered RRC pulse shape. It is shown in \cite{matthe_influence_2014,GFDM_tr} that performance of uncoded GFDM is sensitive to pulse shape choice. Proposed precoded GFDM system will also be sensitive to pulse shape choice as the enhanced noise expressions in (36, 43, 46, 51 and 56) are function of modulation matrix $\bA$ which is a function of pulse shape as given in (7).\\
\indent It can be concluded that precoding schemes are giving good performance over FSFC. BIDFTM precoded GFDM performs best among all precoding schemes.
\subsection{Complexity Computation}\label{sec:result:complexity}
Table~\ref{tab:complexity:tx} compares the complexity of different transmitters and Table~\ref{tab:complexity:rx} compares the complexity of different receivers for two different application scenarios given in \cite{GFDM_tr}. Table~\ref{tab:complexity:app} shows values of $N$ and $M$ considered for different application scenarios. Complexity is computed in terms of number of complex multiplications. Complexities of precoded GFDM systems and uncoded GFDM systems are compared with OFDM system. Spreading factor, $Q=4$, is considered for DFT precoded GFDM. GFDM receiver processing for DFT precoded GFDM is considered to be GFDM-ZF. Number of iterations for DSIC receiver is considered to be 4\cite{datta_gfdm_2012}.
\begin{table}[H]
\center
\begin{tabular}{|c|c|c|}
\hline
& Tactile Internet(TI)& Wireless RAN (WRAN) \\ \hline
$(N,~M)$ & (128, 5) & (16, 127)\\ \hline
\end{tabular}
\caption{Values of $N$ and $M$ for different application scenarios\cite{GFDM_tr}}
\label{tab:complexity:app}
\end{table}
It is observed that the complexity of uncoded GFDM transmitter is about $100$ times higher than OFDM transmitter in case of TI and, around $50$ times greater than OFDM transmitter in case of WRAN. BIDFT precoded and DFT precoded GFDM transmitter has the same order of complexity as uncoded GFDM transmitter. Complexity of SVD precoded GFDM is around two times higher than uncoded GFDM transmitter. Therefore, it can be concluded that there is no significant increase in complexity for BIDFT and DFT precoded transmitter when compared to uncoded GFDM transmitter. However for SVD precoded transmitter complexity is doubled, which is also not a significant increment. \\
\indent GFDM-ZF receiver complexity is much higher than OFDM receiver for instance 50 times in TI scenario. SVD precoded GFDM receiver when SVD is known, two-stage BIDFT precoded receiver and DFT precoded receiver have around same complexity as GFDM-ZF receiver. GFDM-MMSE receiver has very high complexity for example it 2000 times higher than GFDM-ZF in TI scenario. BIDFT precoded receiver complexity, when SVD of channel is not known, is even higher, for instance; it is 100 times higher than the complexity of GFDM-MMSE. It is also important to note that GFDM-DSIC has higher complexity than GFDM-ZF, DFT precoded GFDM and two stage BIDFT precoded GFDM.
\begin{table}[H]
\center
\begin{tabular}{|p{1cm}|p{1.5cm}|p{2.5cm}|p{2.5cm}|p{2.5cm}|p{2.5cm}|}
\hline
CASE& OFDM &Uncoded GFDM & SVD precoded GFDM & BIDFT precoded GFDM & DFT precoded GFDM\\ \hline
TI & $4.4 \times 10^3$ & $4.09 \times 10^5$ & $8.19 \times 10^5$ & $4.09 \times 10^5$ & $4.12 \times 10^5$\\ \hline
WRAN & $8.1 \times 10^3$ & $4.1 \times 10^5$ & $8.25 \times 10^5$ & $4.12 \times 10^5$ & $4.13 \times 10^5$\\ \hline
\end{tabular}
\caption{Number of Complex Multiplications for different Transmitters}
\label{tab:complexity:tx}
\end{table}
\begin{table}[H]
\begin{tabular}{|p{1cm}|p{1cm}|p{1cm}|p{1cm}|p{1cm}|p{1.5cm}|p{1.5cm}|p{1.5cm}|p{1.5cm}|p{1.5cm}|}
\hline
CASE& OFDM &GFDM-ZF & GFDM-MMSE& GFDM-DSIC& SVD-GFDM Rx (KNOWN SVD)& SVD-GFDM Rx (UNKNOWN SVD)& BIDFT-GFDM JOINT PROCESSING&BIDFT-GFDM TWO STAGE PROCESSING & DFT-GFDM\\ \hline
TI & $8.9 \times 10^3$ & $4.19 \times 10^5$ &$8.8 \times 10^7$ &$3.2\times 10^6$  & $4.09 \times 10^5$ & $6.8 \times 10^9$&$3.7 \times 10^6$ & $5 \times 10^5$ & $4.2 \times 10^5$\\ \hline
WRAN & $1.62 \times 10^4$ & $4.16 \times 10^6$ & $2.8 \times 10^9$ & $3.3 \times 10^7$ & $4.1 \times 10^6$&$2.1 \times 10^{11}$ & $2.4 \times 10^7$ & $4.19 \times 10^6$&$4.16 \times 10^6$\\ \hline
\end{tabular}
\caption{Number of Complex Multiplications for different Receivers}
\label{tab:complexity:rx}
\end{table}
\subsection{PAPR of Precoding Techniques}\label{sec:result:papr}
\begin{figure}[h!]
\includegraphics[width = \linewidth]{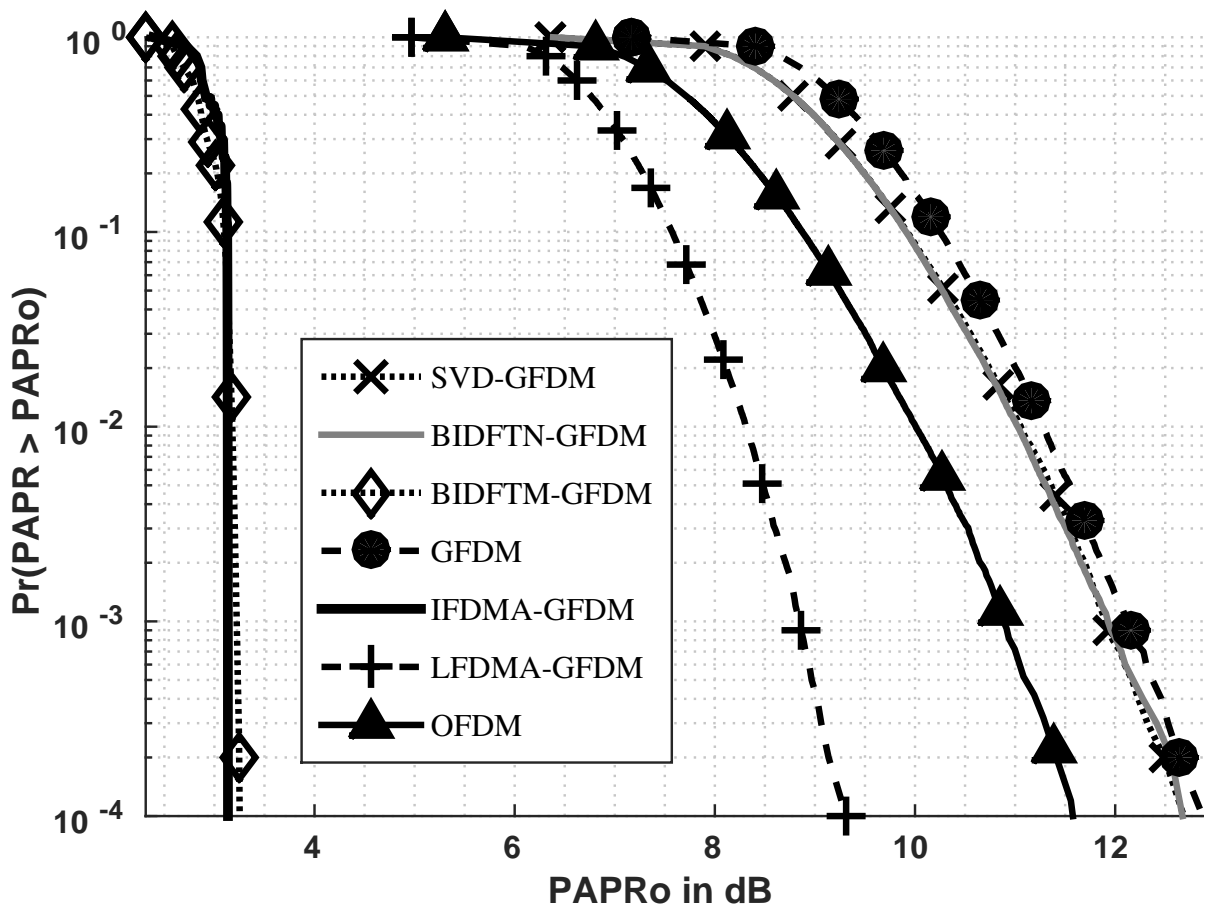}
\caption{Complementary Cumulative Distribution function (empirical) of PAPR for GFDM for $N$=128 and $M$=5, ROF =0.5 (RRC) and Q= 4.}
\label{fig:gfdm:letter:precoding:papr_128x5_coded_gfdm}
\end{figure}

\noindent The impact of precoding on PAPR is presented in Fig.~\ref{fig:gfdm:letter:precoding:papr_128x5_coded_gfdm}. Complementary cumulative distribution function (CCDF) of PAPR is computed using Monte-Carlo simulation. $10^5$ transmitted blocks were generated, where each block has two precoded GFDM symbols. For each precoded GFDM symbol : $N=128$, $M=5$ and ROF for RRC pulse shaping filter = 0.5. For OFDM, $N=128$. Randomly generated data symbols are considered to be QPSK modulated.\\
\indent As expected, GFDM has worse performance than OFDM. Precoding has a positive effect on GFDM. We compare the PAPR value that is exceeded with probability less than $0.1\%$ (Pr\{PAPR $>$ PAPRo $= 10^{-3}\}$). SVD precoded GFDM and BIDFTN precoded GFDM reduces the PAPR by 0.3 dB, but it is still higher than OFDM. DFT precoded GFDM with LFDMA subcarrier mapping reduces PAPR by 3.4 dB and is lower than OFDM. DFT precoded GFDM with IFDMA subcarrier mapping and BIDFTM precoded GFDM reduces PAPR by 9 dB.
\section{Conclusion}  \label{sec:gfdm:ietcomm:precoding:conclusion}
GFDM, a possible waveform for 5G with flexibility to access time--frequency radio resources, is investigated in this work. Two properties of GFDM system are proved (i) product of modulation matrix with its hermitian is a block circulant matrix with circulant blocks and (ii) product of circulant convolution channel matrix with the modulation matrix when multiplied with its hermitian is block circulant matrix. An expression of SINR for MMSE receiver for GFDM is developed using matrix representation for the signal model. The SINR expression is determined in terms of eigenvalues of the product of modulation matrix with its hermitian matrix for AWGN channel and in terms of eigenvalues and eigenvectors of product of circulant convolution channel matrix with the modulation matrix when multiplied with its hermitian for FSFC. It is also shown that the addition of interference and noise values are Gaussian distributed. Using this BER is found. It is found that the BER obtained from simulation and that from the expression developed here match quite well in AWGN channel as well as in FSFC.\\
Three new precoding techniques which improve performance of GFDM system are proposed. The SVD based precoding for GFDM removes interference by orthogonalizing the received symbols. It does not require matrix inversion yet its performance is quite close to that of GFDM-MMSE result. The complexity of the receiver for such precoding is quite high when compared to GFDM-ZF receiver when SVD of channel is not known. However, when SVD of channel is known complexity of the receiver is comparable to GFDM-ZF. Hence, SVD based precoding can be used in cases when channel is constant for multiple transmit instances. Performance of BIDFTN precoding is found to be better than GFDM-ZF and GFDM-DSIC. Two stage BIDFTN precoded GFDM has complexity similar to GFDM-ZF and lesser than GFDM-DSIC as well as gives lower PAPR. Hence, two-stage BIDFTN precoded GFDM should be preferred than GFDM-ZF. BIDFTM precoded as well as DFT precoded GFDM receivers which require complexity similar to GFDM-ZF performs much better than even GFDM-MMSE receiver under FSFC. Apart from this, BIDFTM and DFT precoded GFDM reduces PAPR significantly.
\noindent Precoded GFDM system proposed in this work can give better BER performance than GFDM system with no increase in complexity with the added advantage of decreased PAPR. It can be concluded that BIDFTM precoded GFDM should be preferred to other precoding schemes as it gives better BER performance than other precoded and uncoded receivers and decreases PAPR significantly without any increase in complexity. \\
\bibliographystyle{IEEEtran} 
\bibliography{gfdm}

\end{document}